\documentclass[superscriptaddress]{revtex4}

   \usepackage{calc}
   \usepackage[dvips]{graphicx}
   \usepackage{epsfig}
   \usepackage{amsmath}
   \usepackage{amsthm}
   \usepackage{amssymb}
   \usepackage[boxed]{algorithm}
   \usepackage{algorithmic}
   \usepackage[tight,footnotesize]{subfigure}
   \usepackage[table]{xcolor}
   \usepackage[font=footnotesize,caption=false]{subfig}

\begin{document}

\title{Efficient Collaborative Application Monitoring Scheme for Mobile Networks}

%

\newtheorem{theorem}{Theorem}
\newtheorem{corollary}{Corollary}
\newtheorem{lemma}{Lemma}
\newtheorem{observation}{Observation}
\newtheorem{definition}{Definition}

\newcommand{\remove}[1]{}

\begin{abstract}
New operating systems for mobile devices allow their users to download millions of applications created by various individual programmers, some of
which may be malicious or flawed. In order to detect that an application is malicious, monitoring its operation in a real environment for a significant period of time is often required. Mobile devices have limited computation and power resources and thus are limited in their monitoring capabilities. In this paper we propose an efficient collaborative monitoring scheme that harnesses
the collective resources of many mobile devices, ``vaccinating'' them against potentially unsafe applications. We suggest a new local information flooding algorithm called \emph{TTL Probabilistic Propagation} (TPP). The algorithm is implemented in any mobile device, periodically monitors one or more application and reports its conclusions to a small number of other mobile devices, who then propagate this information
onwards, whereas each message has a predefined TTL. The algorithm is analyzed, and is shown to outperform existing state of the art information propagation algorithms, in terms of convergence time as well as network overhead. The maximal \emph{load} of the algorithm (the fastest arrival rate of new suspicious applications, that can still guarantee complete monitoring), is analytically calculated and shown to be significantly superior compared to any non-collaborative approach.
Finally, we show both analytically and experimentally using real world network data received among others from the \emph{Reality Mining Project}, that implementing the proposed algorithm significantly reduces the number of infected mobile devices.
In addition, we analytically prove that the algorithm is tolerant to several types of Byzantine attacks where some adversarial agents may generate false information, or abuse the algorithm in other way.
\end{abstract}

\maketitle

\section{Introduction}

The market share of Smart-phones is rapidly increasing and is expected to increase even faster with the introduction of $4^{th}$ generation mobile networks, reaching from 350 million in 2009 to one billion by 2012 \cite{Canalys-smartphones}.
Companies that are distributing new mobile devices operating systems had created marketplaces that motivate individuals and other companies to introduce new applications (such as Apple's \emph{App Store} Google's \emph{Android Market}, Nokia's \emph{Ovi Store} and others). The main assumption behind these marketplaces is that users will prefer a mobile device based on an operating system with larger marketplace offerings. It is expected that in the future, various communities will develop additional alternative marketplaces that will not be regulated. These marketplaces will allow mobile users to download from a variety of millions of new applications. An example for such a marketplace is \emph{GetJar}, offering 60,000 downloadable applications for over 2,000 models of mobile devices, counting a total of over 900 million downloads by August 2010 \cite{GetJar}.
The content of most marketplaces is currently not verified by their operators and thus some of the applications they contain may be malicious or contain faulty code segments. Downloading a malicious application from the marketplace is not the only way that a mobile device may be infected by malicious code. This may also happen as a result of a malicious code that manages to exploit a vulnerability in the operating systems and applications or through one of the mobile phone communication channels such as Bluetooth, Wi-Fi, etc' \cite{viruses3-hyponnen,viruses1-kleinberg}.
McAfee's Mobile Security Report for 2008 states that nearly 14\%
of global mobile users have been directly infected or have known someone who was infected by a mobile virus (this number had even increased in the following year) \cite{mcafee1,mcafee2}.
In many cases, in order to detect that an application is malicious, monitoring its operation in a real environment for a significant period of time is required. The monitored data is being processed using advanced machine learning algorithms in order to assess the maliciousness of the application. For a variety of techniques for local monitoring of mobile phone applications see \cite{elovici1-detection-malicious,elovici2-detection-malicious,Anomalous-Mutz,Kim2008}.

In recent years most of the prominent security and privacy threats for communication networks had relied on the use of collaborative methods (e.g., Botnets).
The danger stemming from such threats is expected to significantly increase in the near
future, as argued in \cite{barabasi-science09,viruses1-kleinberg} and others.
The amount of resources a single unit may allocate in order to
defend from such threats without interfering with its routine work is very limited.
Therefore, the development of an efficient collaborative defense infrastructure for mobile users is strongly required \cite{CollaborativeIntrusionDetection-2010}.

\begin{figure}[htb]
   \centering
   \includegraphics[scale=0.25,bb=0 0 1000 650]{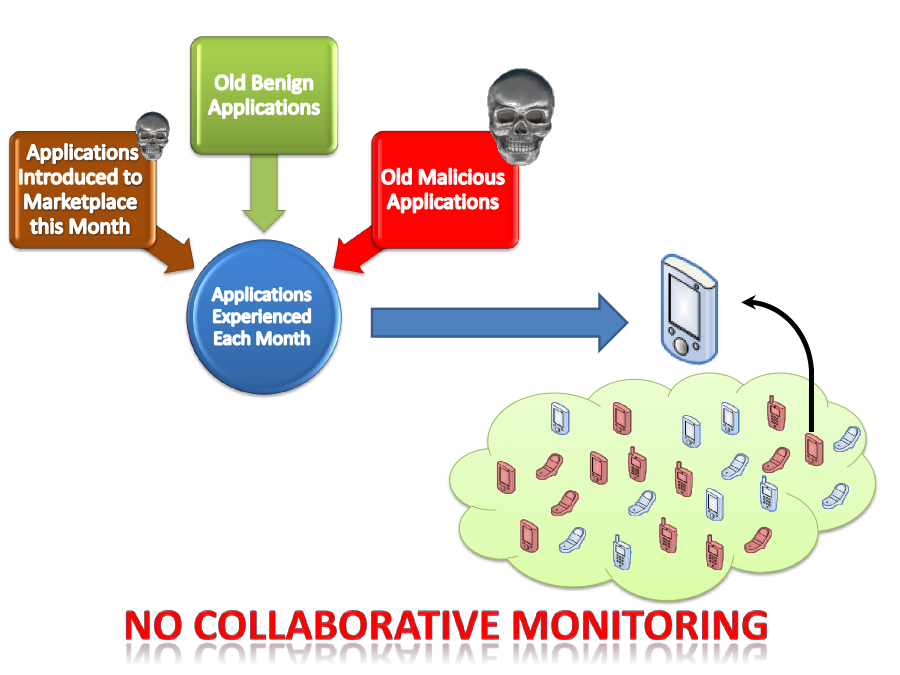}
   \includegraphics[scale=0.25,bb=0 0 800 650]{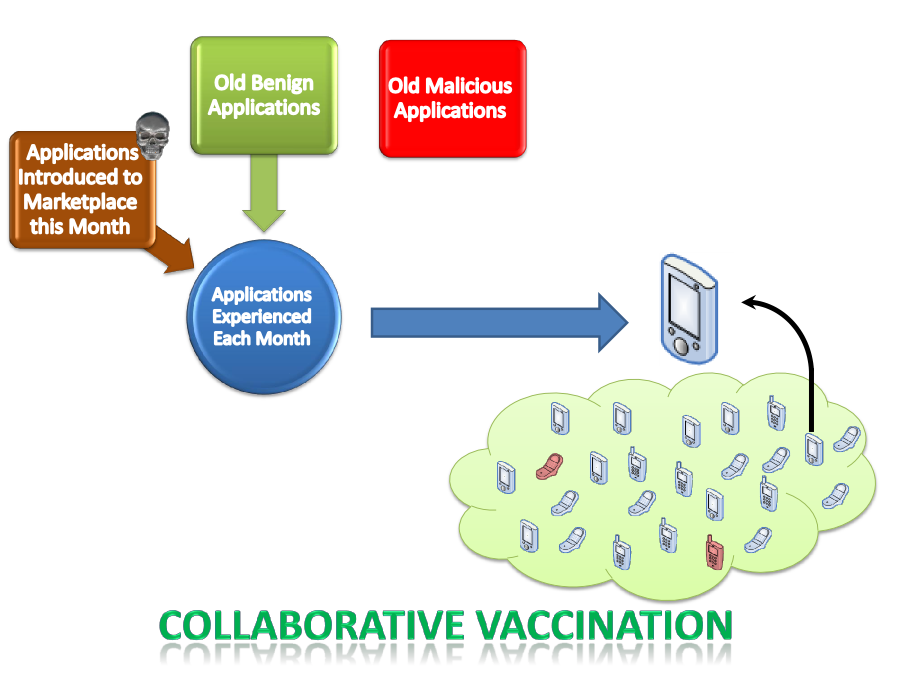}
   \caption{Each month mobile users download and install various kinds of mobile applications. These applications are comprised of ``new applications'' (namely, applications that were only introduced to the marketplace recently), and ``old applications'' (applications that were released to the marketplace in previous months). Whereas new applications may contain a few malicious applications, the vast majority of malicious applications are already in the marketplace for some time. Using the proposed scheme, users of mobile devices become ``vaccinated'' against all of the potentially harmful applications among the old applications. As a result, the number of infected mobile devices decreases dramatically.
   }
   \label{fig.vaccination}
\end{figure}

In this work we propose such a collaborative application monitoring infrastructure, that is capable of dramatically decreasing the susceptibility of mobile devices to malicious applications. Indeed, for every new malicious application that is introduced to the network, there will always be a few users that will encounter it for the first time, and subsequently may be exposed to its negative effects. However, by wisely assimilating this information throughout the network, based on their (poor) experience the vast majority of the mobile devices will be notified on the properties of the new malicious application, well ahead before encountering it. This behavior resembles mammals vaccination system, as a new threat have some (small) probability of damaging the organism (or network), but once defeated, all the cells of the organism are familiar with this threat, and will be able to easily overcome it (or avoid installing it) should they encounter it in the future. As the portion of ``new'' applications (namely, applications that no network user had ever experienced with) is rather small, the mobile devices are kept protected at all times against the vast majority of potentially harmful marketplace applications (this property of the proposed scheme is illustrated in Figure \ref{fig.vaccination}).

This paper presents and analyzes an algorithm that efficiently implements this proposed scheme. The algorithm is shown to provide fast convergence, low network overhead, and is capable of handling a significantly higher number of potential malicious applications concurrently, than any other collaborative or non collaborative approach. In addition, the algorithm is fault tolerance for several known adversarial and Byzantine attacks.
These features of the algorithm are shown both analytically and empirically, using simulation of various random networks, as well as using real life network that are based on MIT's \emph{Reality Mining} project\cite{RealityMining}.

%
In this paper we present a collaborative monitoring algorithm, called \emph{TPP --- TTL Probability Propagation}. The algorithm uses a time to live counter for each alerting message, that is logarithmic in the number of the mobile devices, $n$. Using these short lived messages that are sent between the network's devices the algorithm is analytically shown to provide a prior knowledge of every potential malicious application to the vast majority of the network users. The fastest arrival rate of new applications for which this property can still be maintained (e.g. the \emph{maximal load} of the algorithm) is analytically calculated (see Theorem~\ref{thm.opt.T}), shown to be far superior than any non-collaborative host-based monitoring approach (Corollary \ref{cor.loadratio}) or collaborative approach. Furthermore, we show that the local  cost of the algorithm (namely, the amount of local efforts required from each participating user) monotonically decreases as the network's size increases.

Using real-world numbers, implemented as a service executed by 1,000,000 mobile devices, assuming that each device can locally monitor only a single application per week, and that the requested \emph{vaccination efficiency} is $99\%$ (namely, we allow only $1\%$ of the network's users to be susceptible to attacks), the system is guaranteed to collaboratively monitor 25,566 new applications each month (achieved using an average of 2.763 messages per device per day). A single non-collaborative device under the same requirements would be able to monitor only 5 new applications each month. A similar result was also demonstrated on a real world network, using the data collected in the \emph{Reality Mining Project} \cite{RealityMining} (see Section \ref{sec.experimental}).

In addition, the algorithm is shown to be partially fault tolerant to the presence of Byzantine devices, that are capable of distributing messages containing false information, or selectively refuse to forward some of the messages they receive\footnote{We assume that interference in the messages' content, or generation of messages using false identity are impossible, due to, say, the use of cryptographic means.} (see Theorem \ref{theorem.robust1} in Section \ref{sec.robust} discussing devices who distribute false messages and Corollaries \ref{cor.alg1.time-mute1}, \ref{cor.alg1.time-mute2} and \ref{cor.alg1.cost-mute1} in Section \ref{sec.leeching} concerning devices who block passing messages).

Furthermore, we show that the cost of defending against adversarial abuse of the proposed algorithm grows asymptotically slow whereas the cost of increasing the strength of such attacks grows approximately linearly. This property of the algorithm makes it economically scalable, and desirable to network operators (see more details in Section \ref{sec.robust} and specifically an illustration that appears in  Figure \ref{fig.attack3new}).

The efficiency of the algorithm stems from the generation of an implicit collaboration between a group of random walking agents, released from different sources in the network (and at different times). This technique is novel, as most related works do not take into consideration that the agents might be released collaboratively from different sources \cite{dolev1,dolev2}. Hence, analysis of such systems was limited to the probabilistic analysis of the agents' movements.

%
The rest of this work is organized a follows~:
Related work and comparison to state-of-the-art techniques are presented in Section \ref{sec.related_work}.
The threat model is formally defined in Section \ref{sec.problem}. The \emph{TPP} collaborative algorithm is presented in \ref{sec.analysis}, and its performance is analyzed.  The algorithm's robustness with regards to attacks by adversaries is discussed in Sections \ref{sec.robust} and \ref{sec.leeching}. An alternative model which does not require a random network overlay assumption is discussed in Section \ref{sec.broder}.
A wealth of experimental results are presented in Section \ref{sec.experimental}, where
conclusions and suggestions for future research appears in Section \ref{sec.conc}.
In addition, this paper contains an appendix, discussing in details several issues that are mentioned during the paper.

\section{Related Work}
\label{sec.related_work}
%
Flooding a network with messages intended for a large number of nodes is arguably the simplest form of information dissemination in communication networks (specifically when previous knowledge about the network topology is limited or unavailable). Since the problem of finding the minimum energy transmission scheme for broadcasting a set of messages in a given network is known to be NP-Complete \cite{flooding_NP}, flooding optimization often relies on approximation algorithms. For example, in \cite{flooding_background1}
messages are forwarded according to a set of predefined probabilistic rules, whereas \cite{flooding_background5} discusses a deterministic algorithm, approximating the connected dominating set of each node.

In this work we apply a different approach~---~instead of a probabilistic forwarding of messages, we assign a \emph{TTL} value for each message, guiding the flooding process. The analysis of the algorithm is done by modeling the messages as agents practicing \emph{random walk} in a \emph{random graph} overlay of the network\footnote{An alternative that does not require such an overlay assumption is presented in Section~\ref{sec.broder}.}.
The optimal value of this TTL is shown, guaranteeing a fast completion of the task, while minimizing the overall number of messages sent. The use of a TTL dependent algorithm was previously discussed in \cite{probabilistic-flood-ttl1} (demonstrating $O(\log^{2} n)$ propagation completion time) and \cite{probabilistic-flood-ttl2} (where no analytic bound over the completion time was demonstrated).

\noindent \textbf{Flooding Algorithms}.
The simplest information propagation technique is of course the \emph{flooding} algorithm. It is well known that the basic flooding algorithm, assuming a single source of information, guarantees propagation completion in a worse case cost of $O(n^{2})$ messages ($O(|E|)$ in expanders) 
and time equals to the graph's diameter, which in the case of a random graph $G(n,p)$ equals $O(\frac{\log n}{\log(n \cdot p)}) \approx O(\log n)$
\cite{ER_Graphs,ER_Diameter}. We later show that the \emph{TPP} algorithm discussed in this work achieves in many cases the same completion time, with a lower cost of $O(n \log n)$ messages (Theorems~\ref{corthm.alg1.time} and \ref{corthm.alg1.cost}).

Other flooding algorithms variants use various methods to improve the efficiency of the basic algorithm. One example is a probabilistic forwarding of the messages in order to reduce the overall cost of the propagation process, optionally combined with a mechanism for recognizing and reducing repetitions of packets transitions by the same device \cite{flooding_counterstorm}. Other methods may include area based methods \cite{flooding_counterstorm} or neighborhood knowledge methods \cite{flooding_knowledge2}. 
In many of these works, completion time is traded for a reduced overall cost, which results in a similar cost as
the \emph{TPP} algorithm proposed in this work (namely, $O(n \log n)$), but with a significantly higher
completion time. Additional details on various flooding algorithms can be found in \cite{flooding_survey1}.

An extremely efficient flooding algorithm in terms
of completion time is the network coded flooding algorithm, discussed in \cite{flooding_probability1}.
In this work, dedicated to $G(n,p)$ random graphs, a message is forwarded by any
receiving vertex $\frac{k}{d(v)}$ times, while $k$ is a parameter which depends on
the network's topology \cite{network_coding}. Using this method, the algorithm achieves
a completion time of approximately $O(\frac{n^{3}}{|E|^{2}})$.
%
%
This algorithm, however, is still outperformed by the \emph{TPP} algorithm.

\noindent \textbf{Epidemics Algorithms}.
An alternative approach to be mentioned in this scope is the use of
\emph{epidemic algorithms} \cite{epidemic_book,epidemics-survey}. 
There exist a variety of epidemic algorithms, starting with the basic epidemic
protocol \cite{epidemic_basic1},
through \emph{neighborhood epidemics} \cite{epidemic_neighborhood}
and up to \emph{hierarchical epidemics} \cite{epidemic_hierarchical}.
In general, all the various epidemic variants has a trade-off between number of
messages sent, completion time, and previous knowledge required for the protocols
(concerning the structure of the network). However, the most efficient results
of such approaches are still outperformed by the \emph{TPP} algorithm.

\noindent \textbf{Distributed Coverage Algorithms}.
A different approach for a collaborative assimilation of an important piece of information throughout the network is the use of cooperative exploration algorithms (in either known or unknown environments), guaranteeing that all (or a large enough portion) of the graph is being ``explored'' by agents carrying the alerting messages. Planar networks can be sampled into $\mbox{\bf R}^{2}$, and then be collaboratively explored by a decentralized group of myopic agents
(see swarm coverage examples in
\cite{rekleitis06,UAV-ROBOTICA}).

In \cite{Svennebring1} 
a swarm of ant-like robots is used for repeatedly covering an unknown area, using a real time search method called \emph{node counting}. Using this method, the agents are analytically shown to cover the network graph efficiently.
Another algorithm to be mentioned in this scope is the \emph{LRTA*} algorithm \cite{LRTASTAR}, that was shown  in \cite{Koenig2} to guarantee
cover time of $O(n^2)$ in degree bounded undirected planar graphs. Interestingly, in such graphs, the random walk algorithm is also known to require at most $O(n^{2})$ time (and at least $\Omega(n (\log n)^{2})$) \cite{schramm-covertime-planar}.

In \cite{CC08} a collaborative algorithm that guarantees coverage of regions in the $\mbox{\bf Z}^{2}$ grid, in $O(n^{1.5})$ time, using extremely simple and myopic ``ant like'' mobile agents and using no direct communication by the agents is discussed. In \cite{ICINCO2009} this algorithm was later shown to guarantee (under some limitations) coverage of dynamically expanding domains in $O(n^{2.5} \log n)$ time.

\noindent \textbf{Summary}.
The maximal arrival rate of new applications for which a successful decentralized monitoring can still be guaranteed (namely, the \emph{maximal load}), of the \emph{TPP} algorithm is analyzed in Section \ref{sec.capacity}.

As previously discussed, the efficiency of the \emph{TPP} algorithm is derived from the fact that participating devices form a collaborative infrastructure, using alerting messages that can be sent between each two users --- which implicitly assumes the existence of an appropriate network overlay. The algorithm can still be implemented for any given network topology, assuming no network overlays. This variant of the algorithm is discussed in Section \ref{sec.broder}.
It should also be mentioned that as the \emph{TPP} algorithm uses random elements, it requires approximately $O(\ln^{2} n)$ random bits by each device for its proper execution.

\section{The Collaborative Application Monitoring Problem}
\label{sec.problem}

%
Given a mobile network of $n$ devices, let us denote the network's devices by $V = \{v_{1}, v_{2}, \ldots, v_{n} \}$. Note that the network's topology may be dynamic\footnote{This will later come into effect when messages will be sent between the network's members, at which case the selection of ``an arbitrary network member'' can be assumed to be purely random. Notice that assuming pure randomness among devices selection is not mandatory~---~see Section~\ref{sec.broder} for analysis of the \emph{TPP} algorithm assuming no such network overlays.}.
Each device may occasionally visit the marketplace, having access to $N$ new downloadable applications every month. We assume that downloading of applications is done independently, namely --- that the probability that a user downloads application $a_{1}$ and the probability that the same user downloads application $a_{2}$ are uncorrelated. Let us denote the average number of applications each user downloads each month by $\eta$, out of which $\eta_{n}$ are applications that were introduced to the marketplace this month (``new applications'') and $\eta_{o}$ are applications that were introduced to the marketplace in previous months (``old applications''), such that $\eta_{o} + \eta_{n} = \eta$.
%


For some malicious application $a_{i}$, let $p_{a_{i}}$ denote the application's \emph{penetration probability}~---~the probability that given some arbitrary device $v$, it is unaware of the maliciousness of $a_{i}$. The penetration probability of every new malicious application is 1. Our goal is to verify that at the end of the month, the penetration probability of all malicious applications released during this month are lower than a \emph{penetration threshold} $p_{MAX}$, resulting in a ``vaccination'' of the network with regards to these applications.
This way, although a new malicious application may infect a handful of devices --- the first one it encounters --- the rest of the network would quickly become immune to it.
Formally, for some small $\epsilon$ we require that~:
\[\forall \emph{Malicious application } a_{i} \quad Prob \left( p_{a_{i}} > p_{MAX} \right)  < \epsilon  \]

The rational behind the use of the threshold $p_{MAX}$ is increasing the efficiency of the collaborative system, defending against dominant malicious applications. Given additional available resources, the parameter $p_{MAX}$ can be decreased, resulting in a tighter defense grid (traded for decreased load and increased messages overhead).

We assume that any device $v$ can send a message of some short content to any other device $u$. In addition, we assume that at the initialization phase of the algorithm each device is given a list containing the addresses of some $X$ random network members. This can be implemented either by the network operator, or by distributively constructing and maintaining a random network overlay.

A revised model that does not require this assumption is analyzed in Section \ref{sec.broder}.

We assume that each device can locally monitor applications that are installed on it (as discussed for example in \cite{elovici1-detection-malicious,apap-detecting-malicious}). However, this process is assumed to be  expensive (in terms of the device's battery), and should therefore be executed as fewer times as possible.
The result of an application monitoring process is a non-deterministic boolean value : $\{true, false \}$.
%

False-positive and false-negative error rates are denoted as~:

\[
     P(\emph{Monitoring}(a_{i}) =  \emph{true}  |  A_{i} \emph{ is not malicious})  = E_{+}
     \]
     \[
     P(\emph{Monitoring}(a_{i}) =  \emph{false} |  A_{i} \emph{ is malicious})      = E_{-}
\]

We assume that the monitoring algorithm is calibrated in such way that $E_{+} \approx 0$.

As we rely on the propagation of information concerning the maliciousness of applications, our system might be abused by injection of inaccurate data. This may be the result of a deliberate attack, aimed for ``framing'' a benign application (either as a direct attack against a competitive application, or as a more general attempt for undermining the system's reliability), or simply as a consequence of a false-positive result of the monitoring function.
Therefore, in order for a device $v$ to classify an application $a_{i}$ as malicious, one of the following must hold~:
\begin{itemize}
  \item Device $v$ had monitored $a_{i}$ and found it to be malicious.
  \item Device $v$ had received at least $\rho$ alerts concerning $a_{i}$ from different sources (for some \emph{decision threshold} $\rho$).
\end{itemize}

In addition, note that the information passed between the devices concerns only applications discovered to be malicious. Namely, when an application is found to be benign, no message concerning this is generated. This is important not only for preserving a low messages overhead of the algorithm but also to prevent malicious applications from displaying a normative behavior for a given period of time, after which they initiate their malicious properties. In such cases, soon after an application exits its ``dormant'' stage, it will be detected and subsequently reported, generating a ``vaccination reaction'' throughout the network.

\section{Algorithm, Correctness \& Analysis}
\label{sec.analysis}

We shall now present the \emph{TPP} algorithm. The algorithm is executed by each device separately and asynchronously, where no supervised or hierarchical allocation of tasks, as well as no shared memory are required. Following is a list of the main notations used in the presentation and analysis of the proposed algorithm.

\begin{description}
  \item[$n$] The number of devices participating in the \emph{TPP} algorithm
  \item[$\eta$] Average number of applications each user downloads each month
  \item[$\eta_{n}$] Average number of ``new applications'' users download each month
  \item[$\eta_{o}$] Average number of ``old applications'' users download each month
  \item[$X$] The number of alert messages generated and sent upon the discovery of a malicious application
  \item[$p_{N}$] The ratio $\frac{X}{n}$
  \item[$p_{MAX}$] Penetration threshold --- the maximal percentage of non-vaccinated devices allowed by the network operator
  \item[$E_{-}$] False negative error rate of the local monitoring mechanism
  \item[$T$] Delay time between each two consecutive local monitoring processes
  \item[$N$] Number of new applications introduced to the network each month
  \item[$p_{a_{i}}$] The penetration probability of application $a_{i}$

  \item[$1 - \epsilon$] Confidence level of the correctness of the convergence time estimation
  \item[$\alpha$]  The polynomial confidence level $\ln_{n} \epsilon^{-1}$
  \item[$\lambda_{A}$] Applications arrival rate (number of new applications per time unit)
  \item[$\hat{\lambda}_{A}$]  Maximal applications arrival rate that a non-collaborative device can monitor
  \item[$\lambda_{M}$]  Monitoring rate (number of applications monitored per time unit)
  \item[$\lambda_{T}$]  Number of algorithm's time steps per time unit
  \item[$timeout$] The Time-To-Live counter assigned to alert messages
  \item[$\rho$]  Decision threshold --- the number of alerts a device must receive in order to classify an application as malicious
  \item[$C_{S}$]  Cost of sending a single message
  \item[$C_{M}$]  Cost of locally monitoring a single application
\end{description}

\subsection{\emph{TTL Probabilistic Propagation (TPP)} --- a Collaborative Monitoring Algorithm}
\label{sec.algorithm}

The \emph{TPP} algorithm conceptually relies on the fact that in order to ``vaccinate'' a network with regards to malicious applications, it is enough that only a small number of devices will monitor this application. This way, although the application monitoring process is relatively expensive (in terms of battery and CPU resources), the amortized cost of monitoring each malicious application is kept to a minimum. A detailed implementation of the \emph{TPP} algorithm appears in Algorithm \ref{alg1}.

At its initialization (lines \ref{alg1.line.initstart} through \ref{alg1.line.initend}), all the applications  installed on the device are added to a list of \emph{suspected applications}. In addition, an empty list of \emph{known malicious applications} is created. Once an application is determined as \emph{malicious}, it is added to the known malicious applications list. In case this application was also in the suspected applications list (namely, it is installed on the device, but has not been monitored yet), it is deleted from that list. Once a new application is encountered it is compared to the known malicious applications list, and if found, an alert is sent to the user (alternatively, the application can be chosen to be uninstalled automatically). This feature resembles the long-term memory of the immune system in living organisms. If the new application is not yet known to be malicious, the application is added to the suspected applications list.

Once executed, a periodic selection of an arbitrary application from the list of suspected applications is done, once every $T$ steps (lines \ref{alg1.line.mainstart} through \ref{alg1.line.mainend}). The selected application is then monitored for a given period of time, in order to discover whether it is of malicious properties (see details about such monitoring in Section \ref{sec.problem}). If the application is found to be malicious, it is removed from the list of suspected applications and added to the known malicious applications list (lines \ref{alg1.line.mainremove} and \ref{alg1.line.mainadd2}). In addition, an appropriate alert is produced and later sent to $X$ random devices. The alert message is also assigned a specific TTL value (lines \ref{alg1.line.mainalert1} and \ref{alg1.line.mainalert2}).
Once a network device receives such an alert message it automatically forwards it to one arbitrary device, while decreasing the value of TTL by 1. Once TTL reaches zero, the forwarding process of this message stops (lines \ref{alg1.line.forward1} and \ref{alg1.line.forward2}).
In case a monitored application displays no malicious properties, it is still kept in the list of suspicious applications, for future arbitrary inspections.

A device may also classify an application as malicious as a result of receiving an alert message concerning this application (lines \ref{alg1.line.rhobegin} through \ref{alg1.line.rhoend}). In order to protect begins applications from being ``framed'' (reported as being malicious by adversaries abusing the vaccination system), a device classifies an application as malicious only after it receives at least $\rho$ messages concerning it, from different sources (for a pre-defined \emph{decision threshold} $\rho$).
Note that when a device $v$ receives an alert message concerning application $a_{i}$, it still forwards this message (assuming that $TTL > 0$), even when $v$ has not yet classified $a_{i}$ as malicious (for example, when the number of alert messages received is still smaller than $\rho$).
When the $\rho$-th alerting message concerning an application is received, the device adds the application to its known malicious applications list and removes it from the suspected applications list if needed. The selection of the optimal value of $\rho$ is discussed in Section~\ref{sec.robust}.
The values of $T$, $\rho$ and $TTL$, as well as the number of generated alert messages can be determined by the network operators, or be changed from time to time according to the (known or estimated) values of $n$ and $N$, and the penetration threshold $p_{MAX}$. Selecting an optimal value for $TTL$ is discussed in Section~\ref{sec.analysis}.

\begin{algorithm}
\caption{\emph{TTL Probabilistic Propagation}}
\label{alg1}
\algsetup{indent=0.8cm}
\begin{algorithmic}[1]
\begin{scriptsize}
\STATE \textbf{Initialization} : \label{alg1.line.initstart}
\STATE $\quad$ Let $A(v)$ be the list of installed applications
\STATE $\quad$ Let $\dot{A}(v)$ be the list of suspected applications
\STATE $\quad$ Let $\widetilde{A}(v)$ be the list containing known malicious applications
\STATE $\quad$ Initialize $\dot{A}(v) \leftarrow A(v)$
\STATE $\quad$ Initialize $\widetilde{A}(v) \leftarrow \emptyset$ \\ $\ $\\ \label{alg1.line.initend}

\STATE \textbf{Interrupt} upon encountering a new application $a_{i}$ :
\STATE $\quad$ $\dot{A}(v) \leftarrow \dot{A}(v) \cup \{a_{i}\}$
\STATE $\quad$ \textbf{If} $a_{i} \in \widetilde{A}(v)$ then
\STATE $\quad \quad$ $\dot{A}(v) \leftarrow \dot{A}(v) \setminus \{a_{i}\}$
\STATE $\quad \quad$ Issue an alert to the user concerning $a_{i}$
\STATE $\quad$ \textbf{End if} \\ $\ $ \\

\STATE \textbf{Interrupt} receives malicious application $a_{i}$ notice, for the $j$-th time :
\STATE $\quad$ \textbf{If} $j \geq \rho$ then \label{alg1.line.rhobegin}
\STATE $\quad$ $\quad$ $\dot{A}(v) \leftarrow \dot{A}(v) \setminus \{a_{i}\}$
\STATE $\quad$ $\quad$ $\widetilde{A}(v) \leftarrow \widetilde{A}(v) \cup \{a_{i}\}$
\STATE $\quad$ $\quad$ \textbf{If} $a_{i} \in A(v)$ then
\STATE $\quad  $\quad$ \quad$ Issue an alert to the user concerning $a_{i}$
\STATE $\quad$ $\quad$ \textbf{End if}
\STATE $\quad$ \textbf{End if} \label{alg1.line.rhoend}
\STATE $\quad$ Decrease $TTL$ of report by 1  \label{alg1.line.forward1}
\STATE $\quad$ Forward report to a random network member \\ $\ $ \\ \label{alg1.line.forward2}

\STATE \textbf{Execute every $T$ time-steps} : \label{alg1.line.mainstart}
\STATE $\quad$ Select a random application $a_{i}$ from $\dot{A}(v)$ for monitoring
\STATE $\quad$ \textbf{If} $a_{i}$ is found to be malicious then
\STATE $\quad \quad$ Issue an alert to the user concerning $a_{i}$
\STATE $\quad \quad$ $\widetilde{A}(v) \leftarrow \widetilde{A}(v) \cup \{a_{i}\}$  \label{alg1.line.mainadd2}
\STATE $\quad \quad$ $\dot{A}(v) \leftarrow \dot{A}(v) \setminus \{a_{i}\}$ \label{alg1.line.mainremove}
\STATE $\quad \quad$ Report $a_{i}$ to $X$ random network members  \label{alg1.line.mainalert1}
\STATE $\quad \quad$ Set $TTL = timeout$   \label{alg1.line.mainalert2}
\STATE $\quad$ \textbf{End if} \label{alg1.line.mainend}
\end{scriptsize}
\end{algorithmic}
\end{algorithm}

\subsection{Optimal Parameters for Guaranteeing Successful Monitoring}

\noindent \textbf{Outline of Analysis}.
In order to analyze the algorithm's behavior, we shall model the movements of the notification messages between the network's devices as random walking agents, traveling in a random graph $G(n,p)$. Taking into account the fact that the messages have limited lifespan (namely, $TTL$), a relation between the size of the graph and the lifespan of the agents is produced. Having the value of $TTL$ that guarantees a coverage of the graph, the algorithm's completion time, as well as the overall number of messages sent, can then be calculated.

While analyzing the performance of the \emph{TPP} algorithm we imagine a directed \emph{Erd\"{o}s-Renyi} random graph $G(V,E) \sim G(n,p_{N})$, where $p_{N} = \frac{X}{n}$.
The graph's vertices $V$ denote the network's devices, and the graph's edges $E$ represent messages forwarding connections. Notice that as $G$ is a random graph, it can be used for the analysis of the performance of the \emph{TPP} algorithm, although the message forwarding connections of the \emph{TPP} algorithm are dynamic (notice again that an alternative model which does not require this randomness assumption is discussed in Section~\ref{sec.broder}). In addition, although the identity of the ``neighbors'' of a vertex $v$ in the real network overlay may change from time to time (as the overlay graph can be dynamic), it can still be modeled by static selection of $X$ random neighbors of $v$.

Observing some malicious application $a_{i}$, every once in a while some device which $a_{i}$ is installed on randomly selects it for monitoring. With probability $(1 - E_{-})$ the device discovers that $a_{i}$ is malicious and issues alerts  to $X$ network's members.
We look at these reports as the system's ``agents'', and are interested in finding~:
\begin{itemize}
  \item The time it takes the graph to be explored by the agents. Namely, the time after which every device was visited by at least $\rho$ agents (and is now immune to $a_{i}$).
  \item Total number of messages sent during this process.
  \item The minimal TTL which guarantees a successful vaccination of the network.
\end{itemize}

Note that the agents have a limited lifespan, equals to $timeout$.
As the graph is a random graph, the location of the devices in which $a_{i}$ is installed is also random. Therefore, as they are the sources of the agents, we can assume that the initial locations of the agents are uniformly and randomly distributed along the vertices of $V$.
In compliance with the instruction of the \emph{TPP} algorithm, the movement of the agents is done according to the random walk algorithm.

Application $a_{i}$ is installed on $n \cdot p_{a_{i}}$ devices, each of which monitors a new application every $T$ time steps. Upon selecting a new application to monitor, the probability that such a device will select $a_{i}$ is $\frac{1}{N}$. The probability that upon monitoring $a_{i}$ the device will find it to be malicious is $(1 - E_{-})$, in which case it will generate $n \cdot p_{N}$ alerting messages.
The expected number of new agents created at time $t$, denoted as $\hat{k}(t)$, therefore equals~:
\[\hat{k}(t) = \frac{n^{2} \cdot p_{a_{i}} \cdot p_{N}}{T \cdot N} (1 - E_{-})\]
and the accumulated number of agents which have been generated in a period of $t$ time-steps is therefore
$k_{t} = \sum_{i \leq t} \hat{k}(i)$.

The value of $timeout$ (the assigned TTL) is selected in such a way that the complete coverage of the graph, and therefore its vaccination against $a_{i}$, is guaranteed (in probability greater than $1 - \epsilon$). We now artificially divide the mission to two phases, the first containing the generation of agents and the second discussing the coverage of the graph. Note that this division ignores the activity of the agents created in the second phase. Note also that the fact that the agents are working in different times (and in fact, some agents may already vanish while others have not even been generated yet) is immaterial. 
The purpose of this technique is to ease the analysis of the flow of the vaccination process.

Denoting the completion time by $T_{Vac}$ we therefore have~:
\[T_{Vac} \leq T_{Generation} + T_{Propagation}\]

It is easy to see that essentially $T_{Propagation} \triangleq timeout$. We now artificially set~:
\[timeout = \lambda \cdot (T_{Generation} + timeout)\]
\[T_{Generation} = (1 - \lambda)  \cdot  (T_{Generation} + timeout)\]

From this we can see that~:
\[
    T_{Generation} = \frac{(1 - \lambda)}{\lambda} \cdot timeout
\]

We later demonstrate an upper bound for $timeout$, based on the number of agents created in $t \leq T_{Generation}$ (ignoring the activity of the agents created between $t = T_{Generation}$ and $t = T_{Generation} + timeout$).

We now examine the case of $\lambda = 0.5$ (which we show to be the optimal selection for $\lambda$, in the paper's Appendix). 
In this case, we can now write:
$T_{Vac} \leq 2 \cdot timeout$.

Let us denote the number of agents created in the first $T_{Generation}$ time-steps by $k = k_{T_{Generation}}$. We now find the time it takes those $k$ agents to completely cover the graph $G$, and from this, derive the value of $timeout$.

Let us denote by \emph{$\rho$-coverage} of a graph the process the result of which is that every vertex in the graph was visited by some agent at least $\rho$ times.

\begin{theorem}
\label{theorem.m}
The time it takes $k$ random walkers to complete a $\rho$-coverage of $G$ in probability greater than $1-\epsilon$ (denoted as T(n)) can be bounded as follows~:
\begin{displaymath}
\frac{2 \left(\rho -\ln \frac{\epsilon}{n} \right)}{1-e^{-\frac{3 k}{2 n (1 - \frac{1}{\ln n})}}} \leq T(n) \leq \frac{2 \left(\rho -\ln \frac{\epsilon}{n} \right)}{1-e^{-\frac{k}{2n}}}
\end{displaymath}
\begin{proof}
See Appendix for complete proof.
\end{proof}
\end{theorem}

We now show how to select a value of $timeout$ that guarantees a successful vaccination process~:
\begin{theorem}
\label{theorem.timeout1}
For every values of $timeout$ that satisfy the following expression, the \emph{TPP} algorithm is guaranteed to achieve successful vaccination for any penetration threshold $p_{MAX}$, in probability greater than $1 - \epsilon$~:
\[\frac{2 \left(\rho -\ln \frac{\epsilon}{n} \right)}{timeout (1-e^{- timeout \cdot \frac{n \cdot p_{MAX} \cdot p_{N}}{2 T \cdot N} (1 - E_{-})})} = 1\]
\begin{proof}
Recalling the expected number of agents generated at each time step, the expected number of agents $k$ that appears in Theorem \ref{theorem.m} equals~:
%
\[E[k] = \sum_{i \leq T_{Generation}} \frac{n^{2} \cdot p_{a_{i}} \cdot p_{N}}{T \cdot N} (1 - E_{-}) \]

A successful termination of \emph{TPP} means that the penetration probability of (any) malicious application is decreased below the threshold $p_{MAX}$. Until this is achieved, we can therefore assume that this probability never decreases below $p_{MAX}$~:
\[\forall t < T_{Generation} : p_{a_{i}} \geq p_{MAX}\]

Therefore, we can lower bound the number of agents as follows~:
\[k \geq timeout \cdot \frac{n^{2} \cdot p_{MAX} \cdot p_{N}}{T \cdot N} (1 - E_{-})\]
%

Assigning $timeout = m$ into Theorem \ref{theorem.m}, successful vaccination is guaranteed for~:
%
\[timeout = \frac{2 (\rho -\ln \frac{\epsilon}{n} )}{1-e^{-\frac{k}{2n}}} \leq \frac{2 \left(\rho -\ln \frac{\epsilon}{n} \right)}{1-e^{- \frac{n \cdot p_{MAX} \cdot p_{N}}{2 T \cdot N \cdot timeout^{-1}} (1 - E_{-})}} \]
and the rest is implied
\end{proof}
\end{theorem}

\subsection{Number of Messages and Time Required for Vaccination}

From the value of $timeout$ stated in Theorem \ref{theorem.timeout1}, the vaccination time $T_{Vac}$ as well as the overall cost of the \emph{TPP} algorithm can now be extracted. The cost of the algorithm is measured as a combination of the overall number of messages sent during its execution and the total number of monitoring activities performed. Let us denote the cost of sending a single message by $C_{S}$ and the cost of executing a single local monitoring process by $C_{M}$.
\begin{observation}
\label{obs.simple.timeout.results}
For any $timeout = \tau$ which satisfies Theorem \ref{theorem.timeout1}, the time and cost of the \emph{TPP} algorithm can be expressed as~:
\[ T_{Vac} = O\left( \tau \right)  \quad , \quad  M = O(k \cdot \tau \cdot C_{S} + \frac{k}{X} C_{M}) =\]
\[O\left( \frac{p_{MAX} \cdot p_{N}}{n^{-2} T \cdot N} \cdot (1 - E_{-}) \cdot \left( \tau^{2} \cdot C_{S} + \frac{\tau }{n \cdot p_{N}} \cdot C_{M} \right) \right)\]
\end{observation}

\begin{theorem}
\label{corthm.alg1.time}
The time it takes a network that implements the \emph{TPP} algorithm to guarantee vaccination against all the malicious applications from an $N$ available applications is upper bounded as follows~:
\[ T_{Vac} \leq 4 \sqrt{\frac{T \cdot N \left(\rho + (\alpha + 1) \ln n \right)}{n \cdot p_{MAX} \cdot p_{N} \cdot (1 - E_{-})}}
 = O\left( \rho + \ln n + \frac{T \cdot N}{p_{MAX} \cdot (1 - E_{-}) \cdot \ln n} \right)\]
\begin{proof}
Let us assume that $\epsilon$ is polynomial in $\frac{1}{n}$, namely~:
$\epsilon = n^{-\alpha} \ s.t. \ \alpha \in \mathbb{Z}^{+}$.

Using the bound $(1 - x) < e^{-x}$ for $x < 1$ we can see that when assuming %
\footnote{We aspire that the number of messages each device is asked to send upon discovering a new malicious application is kept to a minimum. As the value of $P_{N}$ must be greater than $\frac{\ln n}{n}$ in order to guarantee connectivity \cite{ER_Graphs2}, it is safe to assume $P_{N} = O(\frac{\ln n}{n})$. Note that a connected pseudo-random graph can also be generated with $p_{N} = O(\frac{1}{n})$ \cite{dolev1}.
We lower bound $p_{N}$ as we are interested in demonstrating the result for any random graph $G(n,p)$.
In addition, we later show that $timeout \approx O(\log n)$. It is also safe to assume that $N \approx \Omega(\ln n)$ and that $P_{MAX} \approx O\left(\frac{1}{\ln n}\right)$. This assumption in later discussed in great details.} %
that~:
\[timeout \cdot \frac{n \cdot p_{MAX} \cdot p_{N}}{2 T \cdot N} (1 - E_{-}) < 1\]
Theorem \ref{theorem.timeout1} can be written as~:
\[\rho + (\alpha + 1) \ln n \geq timeout^{2} \cdot \frac{n \cdot p_{MAX} \cdot p_{N} \cdot (1 - E_{-})}{4 T \cdot N}\]
and therefore~:
\[ timeout \leq \sqrt{\frac{4 T \cdot N \left(\rho + (\alpha + 1) \ln n \right)}{n \cdot p_{MAX} \cdot p_{N} \cdot (1 - E_{-})}} \]

Assigning this approximation of $timeout$ into the assumption above, yields~:
\[ \sqrt{\frac{2 T \cdot N}{n \cdot p_{MAX} \cdot p_{N} (1 - E_{-})} \cdot 2 \left(\rho + (\alpha + 1) \ln n \right)}
< \frac{2 T \cdot N}{n \cdot p_{MAX} \cdot p_{N} (1 - E_{-})}\]
which is satisfied for the following values of $p_{N}$~:
\[ p_{N}  < \frac{T \cdot N}{n \cdot p_{MAX} \cdot (\rho + (\alpha + 1) \ln n) (1 - E_{-})}\]
For constant (or smaller) values of $p_{MAX}$ and number of applications larger than $O(\ln n)$ we can safely assign $p_{N} = \frac{\ln n}{n}$.
As the vaccination time equals twice the value of $timeout$, the rest is implied.

Assigning the upper bound for $p_{N}$ into Theorem \ref{corthm.alg1.time} immediately yields $O(\rho + \ln n)$. When assigning $\frac{\ln n}{n}$ as a lower bound for $p_{N}$ which guarantees connectivity \cite{ER_Diameter}, the following expression is received~:
\[ T_{Vac} \leq 4 \sqrt{ \frac{T \cdot N \cdot \left(\rho + (\alpha + 1) \ln n \right)}{\ln n \cdot p_{MAX} \cdot (1 - E_{-})} }\]
However, using the upper bound for $p_{N}$ derived above, we see that~:
\[ \frac{\ln n}{n}  < \frac{T \cdot N}{n \cdot p_{MAX} \cdot (\rho + (\alpha + 1) \ln n) (1 - E_{-})}\]
which in turn implies that~:
\[ \rho + (\alpha + 1) \ln n <  \frac{T \cdot N}{p_{MAX} \cdot(1 - E_{-}) \cdot \ln n } \]
Combining the two yields~:
\[ T_{Vac} = O\left( \frac{T \cdot N }{p_{MAX} \cdot (1 - E_{-}) \cdot \ln n }  \right) \]
Note that although $O \left(\frac{T \cdot N}{p_{MAX} (1 - E_{-})} \right)$ is allegedly independent of $n$, assigning the connectivity lower bound $P_{N} > \frac{\ln n}{n}$ into the upper bound for $p_{N}$ we can see that :
\[\frac{T \cdot N}{p_{MAX} (1 - E_{-})} = \Omega(\rho \ln n + \ln^{2} n)\]
\end{proof}
\end{theorem}

\begin{theorem}
\label{corthm.alg1.cost}
The overall cost (messages sent and monitoring) of a network that implements the \emph{TPP} algorithm for guaranteeing vaccination against all the malicious applications from an $N$ available applications is upper bounded as follows~:
\[ M \leq k \cdot timeout \cdot C_{S} + \frac{k}{X} \cdot C_{M} \leq \]
\[ \leq 4 n \left(\rho + (\alpha + 1) \ln n \right) C_{S} + 2 C_{M}    \sqrt{\frac{ n \left(\rho + (\alpha + 1) \ln n \right) \cdot p_{MAX} \cdot (1 - E_{-})}{p_{N} \cdot T \cdot N }} = \]
\[ O\left( (n \rho + n \ln n) C_{S} +  \left(\frac{n}{\ln n} +  n (\rho + \ln n) \frac{p_{MAX} (1 - E_{-})}{T \cdot N} \right) C_{M} \right)\]
\begin{proof}
When the vaccination process is completed, no new messages concerning the malicious application are sent.
The above is received by assigning the approximated value of $timeout$ into Observation \ref{obs.simple.timeout.results}.
\end{proof}
\end{theorem}

In networks of $E_{-} < 1 - o(1)$, provided that\footnote{See Section~\ref{sec.robust} for more details.} $\rho = O(\ln n)$, and remembering that in this case $\frac{T \cdot N}{p_{MAX} (1 - E_{-})} = \Omega(\ln^{2} n)$ Theorem~\ref{corthm.alg1.cost} is dominated by~:
\[ M = O\left( n \ln n C_{S} +  \frac{ n}{\ln n} C_{M} \right)\]

\section{Maximal Load of the \emph{TPP} Algorithm}
\label{sec.capacity}

We shall now calculate the maximal load of the \emph{TPP} algorithm, namely, the maximal rate at which new applications can be introduced into the system, while guaranteeing that each of them would still be monitored by enough devices.

\begin{definition}
Let $\lambda_{A}$ denote the rate at which new applications are introduced to the system (units are \#applications per time units).
\end{definition}

\begin{definition}
Let $\lambda_{M}$ denote the rate at which each device monitors an arbitrary new application (units are \#applications per time units).
\end{definition}

\begin{definition}
Let $\lambda_{T}$ denote the rate at which alert messages are processed throughout the system (namely, how many ``time steps'' pass per time unit).
\end{definition}

We can now derive the maximal load of the system, as follows~:
\begin{theorem}
\label{thm.opt.T}
An upper bound for the maximal applications load $\lambda_{A}$ is~:
\[
\lambda_{A} \leq  \frac{n \cdot p_{MAX} \cdot p_{N} \cdot (1 - E_{-})}{16\left(\rho + (\alpha + 1) \ln n \right)}  \cdot \lambda_{\widetilde{T}}^{2}  \cdot  \lambda_{M}
\]
where:
\[
\lambda_{\widetilde{T}} \triangleq \min \left\{ \lambda_{T} , \frac{2\rho + 2(\alpha+1)\ln n}{1 - e^{-\frac{3}{2} \frac{p_{N}}{(1 - \frac{1}{\ln n}) (1 - p_{N} - \frac{1}{n})} }} \right\}
\]
\begin{proof}

Interpreting $\lambda_{T}^{-1}$ as the number of months it takes to process a single step of the algorithm, we can now assign $N = \lambda_{A}$ and $T = \lambda_{M}^{-1}$ into Theorem \ref{corthm.alg1.time}, requesting that the upper bound over $T_{Vac}$ shall be kept smaller than 1.
$\lambda_{\widetilde{T}}$ denotes the maximal number of steps that may be required by the algorithm to complete the proliferation of the alerting messages produced in the last monitoring of each month, calculated using the lower bound part of Theorem~\ref{theorem.m}.\footnote{The number of nodes to be covered that is assigned to the expression of Theorem~\ref{theorem.m} is $n - n \cdot p_{N} - 1$ as the $n \cdot p_{N}$ alerting messages are known to be sent to different nodes, at the first step of the algorithm. In addition, the node that generates these messages is naturally aware of their content.}
\end{proof}
\end{theorem}

\begin{observation}\label{obs.load1}
For $\rho = O(\ln n)$ and constant values of $p_{MAX}$ and $E_{-}$, then for $p_{N} > \frac{\ln n}{n}$ the maximal load is monotonically increasing with $n$.
\end{observation}

We can now calculate the algorithm's \emph{benefit factor}, comparing the system's maximal load to the maximal load of any non collaborative monitoring algorithm~:

\begin{corollary}
\label{cor.loadratio}
Let us denote by $\hat{\lambda}_{A}$ the maximal load of a single device. The ratio between the maximal load of applications a single device can non-collaboratively monitor and the maximal load of the \emph{TPP} algorithm is~:
\[
\frac{\lambda_{A}}{\hat{\lambda}_{A}} = \lambda_{\widetilde{T}}^{2} \cdot (1-p_{MAX}) \cdot p_{MAX} \cdot  \frac{n \cdot p_{N}}{16\left(\rho + (\alpha + 1) \ln n \right)}
\]
\begin{proof}
We can easily see that~:
\[
\hat{\lambda}_{A} = (1 - E_{-})  \cdot  \frac{1}{1 - p_{MAX}}  \cdot  \lambda_{M}
\]
Dividing $\lambda_{A}$ by $\hat{\lambda}_{A}$ the rest is implied.
\end{proof}
\end{corollary}

\begin{observation}\label{obs.load2}
For $\rho = O(\ln n)$ and constant values of $p_{MAX}$ and $E_{-}$, then for $p_{N} > \frac{\ln n}{n}$ and regardless of the monitoring rate $\lambda_{M}$ the \emph{benefit factor} of participating in the collaborative \emph{TPP} monitoring algorithm is asymptotically greater than 1, and is monotonically increasing with $n$.
\end{observation}

\noindent \textbf{Numeric Example I (\emph{Large Network})~: }
assume a network of $n = 1,000,000$ devices who participate in the collaborative effort.
We shall require that $p_{MAX} = 0.01$ (namely, that 99\% of the network must be fully vaccinated) and that the reliability of the analysis should be $99.9\%$ (as $n = 1,000,000$ it means that $\alpha = 0.5$).
Disregarding adversarial attacks, we can use $\rho = 1$, and in addition, we shall assume that the error probability is very small. We also assume that the time that is required for a message to be processed and forwarded is approximately 5 seconds (namely, $\lambda_{T} = 720$, assuming time units of hours). Upon discovery of a malicious application, we shall assume that a device sends this information to 400 network members (namely, $p_{N} = \frac{1}{2500}$). We shall also assume that each device can monitor a single application per week (namely, $\lambda_{M} = \frac{1}{24 \cdot 7}$). Using Theorem \ref{thm.opt.T} we see that $\lambda_{A} \leq 35.51163$ which equals to 25,566 new applications per month!\footnote{Notice that by using days (or weeks) instead of hours, the maximal monthly load of the algorithm could have further been increased. However, this would have also increased the time between the initial discovery of a malicious application and time at which the vaccination process is completed. Selecting a short time unit (i.e. an hour) ensures that this time is kept to a minimum.}

As to the number of messages each participating device is required to send, using Theorem \ref{corthm.alg1.cost} we can see that each device sends on average of at most $6 \cdot \ln n$ messages during each month, namely --- 2.763 messages on average per day.
Comparing this to the performance of a single device relying solely on a non collaborative host-based monitoring, such a device could cope with monitoring less then 5 new applications each month, as $\hat{\lambda}_{A} \leq 0.006$). This reflects a benefit factor of more than 5,000 in the maximal load of suspicious applications.

\noindent \textbf{Numeric Example II (\emph{Reality Mining})~: }
In order to assess the performance of the algorithm for real-world networks, we have used the data generated by the \emph{Reality Mining} project \cite{RealityMining}. This project analyzed 330,000 hours of continuous behavioral data logged by the
mobile phones of 94 subjects, forming a mobile based social network. Using this network, we can estimate the efficiency of the \emph{TPP} algorithm, as follows. Based on the \emph{Reality Mining} data we can calculate that $n = 94$, $\alpha = 1.52$ (in order to guarantee a 99.9\% confidence level) and $p_{N} = 0.092123$. In addition, as in the previous example, we assume that each device can monitor a single application per week ($\lambda_{M} = \frac{1}{24 \cdot 7}$) and that the time required for a message to be processed equals 5 seconds ($\lambda_{T} = 720$ and $\lambda_{\widetilde{T}} =  139$) and were no adversaries are present $(\rho = 1)$. Then the maximal load of the system $\lambda_{A}$ as a function of the penetration threshold $p_{MAX}$ equals (using Theorem \ref{thm.opt.T})~:
\[
\lambda_{A} \leq 4.997 \cdot p_{MAX}
\]

\begin{figure}[htb]
   \centering
   \includegraphics[scale=0.5,bb=0 0 900 800]{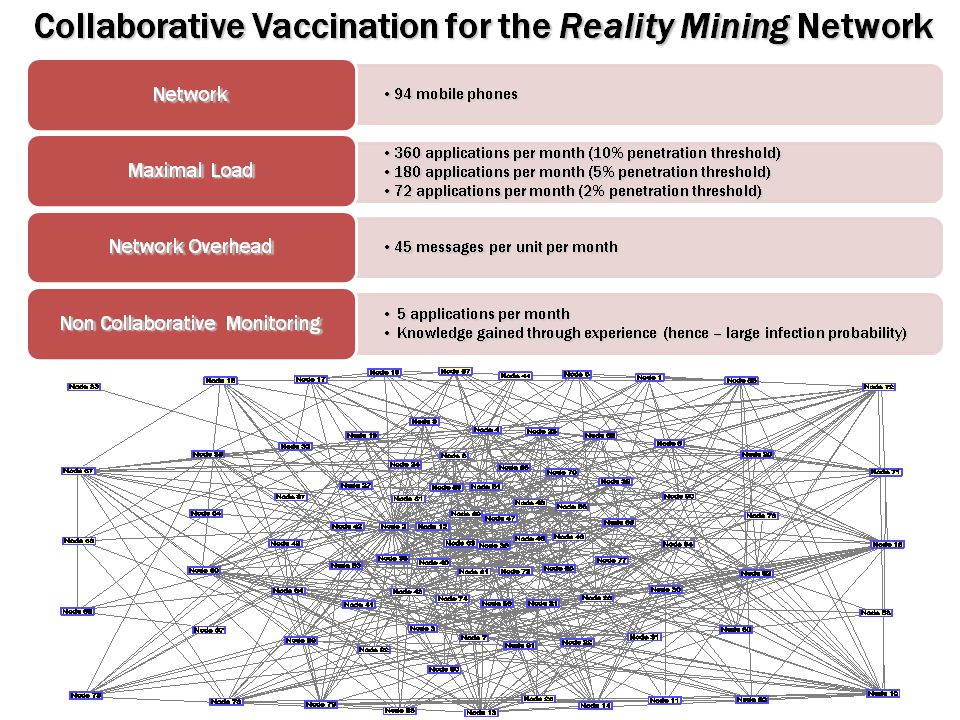}
   \caption{Maximal monthly load of the \emph{TPP} algorithm implemented over the \emph{Reality Mining} network. Note that the maximal load increases as the vaccination mechanism loosens (namely, guarantees vaccination for a slightly smaller portion of the network). For example, where for 98\% vaccination the algorithm guarantees monitoring of 72 new applications per month, when we require that only 90\% of the network's devices are vaccinated, the number of applications that can be monitored each month increases to 360.
   }
   \label{fig.load-RM}
\end{figure}

The maximal load of the algorithm for this network appears in Figure \ref{fig.load-RM}.

It is important to notice that although the number of applications the average user downloads each month is far smaller than 5,000, the importance of the collaborative monitoring scheme is still high, as the main merit of the proposed scheme lays in its predictive ability. Namely, whereas a user downloading 5 applications each month may locally monitor them successfully, should one of them be malicious, the user will fall victim to its malicious behavior. However, should users that participate in the proposed collaborative monitoring scheme try to download a malicious application, they will be \emph{immediately notified} about the maliciousness of the application (in high probability), before they are exposed to its dangers. Indeed, the first users that encounter a new malicious application may fall victim to its perils --- an unavoidable property of any host-based monitoring system. However, using the proposed collaborative scheme this is kept to a minimum.

\section{Avoiding Benign Applications ``Frame'' Through Adversarial Use of the \emph{TPP} Vaccination Algorithm}
\label{sec.robust}

As mentioned in previous sections, the \emph{TPP} algorithm is fault tolerant to the presence of adversarial devices which try to compromise the system's integrity by causing a large enough portion of the network to consider (one or more) benign applications as malicious. This aspect of the algorithm is crucial, as it is a collaborative algorithm which relies on the participation of as many mobile devices as possible~---~devices who should be guaranteed that the efficient collaborative monitoring will not be accompanied with erroneous results.
In the \emph{TPP} algorithm, this fault tolerance is being provided by the introduction of the $\rho$ ``decision threshold'' into the decision system. Namely, the fact that an application is being marked as malicious only when at least $\rho$ relevant messages (from different origins) are being received. Relying on the common agreement of several devices as a tool for overcoming system noise (which can be either coincidental or intentional) is often used in swarm based systems. For example, a similar mechanism called ``threshold cryptography'' is used for enabling the collaborative use of cryptographic tasks (see for example \cite{TresholdCrypto1,TresholdCrypto3}.


\begin{definition}
Let us denote by $P_{Attack} (TTL, \rho, \frac{k}{n}, \varepsilon)$ the probability that a ``framing attack'' done by a group of $k$ organized adversaries will successfully convince at least $\varepsilon \cdot n$ of the network's devices that some benign application $a_{i}$ is malicious.
\end{definition}

A trivial example is the use of very large values for $TTL$, which allow a group of $k$ adversaries to convince the entire network that any given application is malicious, provided that $k > \rho$, namely~:
%
\[
      \forall k\geq 1 \quad , \quad
      \forall \rho \leq k \quad , \quad
      \forall \varepsilon < 1 \quad , \quad
        \lim_{TTL \rightarrow \infty} P_{Attack} (TTL, \rho, \frac{k}{n}, \varepsilon) = 1
\]
\begin{theorem}
\label{theorem.robust1}
The probability that $k$ attackers will be able to make at least an $\epsilon$ portion of the network's devices treat some benign application $a_{i}$ as malicious, using the \emph{TPP} algorithm with a decision threshold $\rho$ is~:
\[
P_{Attack} \left(TTL, \rho, \frac{k}{n}, \varepsilon\right) \leq  1 - \Phi \left( \sqrt{n} \cdot \frac{\varepsilon - \tilde{P}}{\sqrt{\tilde{P} (1-\tilde{P})    }} \right)
\]
where~:
\[
\tilde{P} =
  e^{(\rho - TTL \cdot (1-e^{-\frac{k \cdot p_{N}}{2}}))} \cdot \left( \frac{TTL \cdot (1-e^{-\frac{k \cdot p_{N}}{2}})}{\rho} \right)^{\rho}
\]
and where $\Phi(x)$ is the cumulative normal distribution function, defined as~:
\[
\Phi(x) = \frac{1}{\sqrt{2 \pi}} \int_{-\infty}^{x} e^{-\frac{1}{2}t^{2}} dt
\]
and also provided that~:
\[
\rho > TTL \left(1 - e^{-\frac{k \cdot p_{N}}{2}}\right)
\]
\begin{proof}
We use Lemma~\ref{lemma.prob1} to calculate the probability that a device $v \in V$ will be reported of some malicious application $a_{i}$ by a message sent by one of the $k$ adversaries at the next time-step. This is yet again a Bernoulli trial with~:
\[
p_{s} = 1-e^{-\frac{(k \cdot n \cdot p_{N})}{2n}} = 1-e^{-\frac{k \cdot p_{N}}{2}}
\]

Denoting as $X_v(t)$ the number of times a notification message had arrived to $v$
after $t$ steps, using \emph{Chernoff} bound~:
\[ P[X_v(t) > (1+\delta) t \cdot p_{s}] < \left(\frac{e^{\delta}}{(1+\delta)^{(1+\delta)}}\right)^{t \cdot p_{s}}\]
in which we set $\delta = \frac{\rho}{t \cdot p_{s}} - 1$.
We can therefore see that~:
\[
  \tilde{P} \triangleq P_{Attack} (TTL, \rho, \frac{k}{n}, n^{-1}) = P[X_v(TTL) > \rho ]  <
  e^{(\rho - TTL \cdot p_{s})} \cdot \left( \frac{TTL \cdot p_{s}}{\rho} \right)^{\rho}
\]

It is important to note that the \emph{Chernoff} bounds requires that $\delta > 0$. This is immediately translated to the following requirement~:
\[
\rho > TTL (1 - e^{-\frac{k \cdot p_{N}}{2}})
\]

As we want to bound the probability that at least $\varepsilon n$ of the devices are deceived, we shall use the above as a success probability of a second \emph{Bernoulli} trial. As $n$ is large, the number of deceived devices can be approximated using normal distribution~:
\[
P_{Attack} \left(TTL, \rho, \frac{k}{n}, \varepsilon\right) \leq  1 - \Phi \left( \frac{\varepsilon \cdot n - n \cdot \tilde{P}}{\sqrt{n \cdot \tilde{P} (1-\tilde{P})    }} \right)
\]
and the rest is implied.
\end{proof}
\end{theorem}

\begin{figure}[htbp]
\begin{center}
\includegraphics[scale=0.27,bb=0 0 800 650]{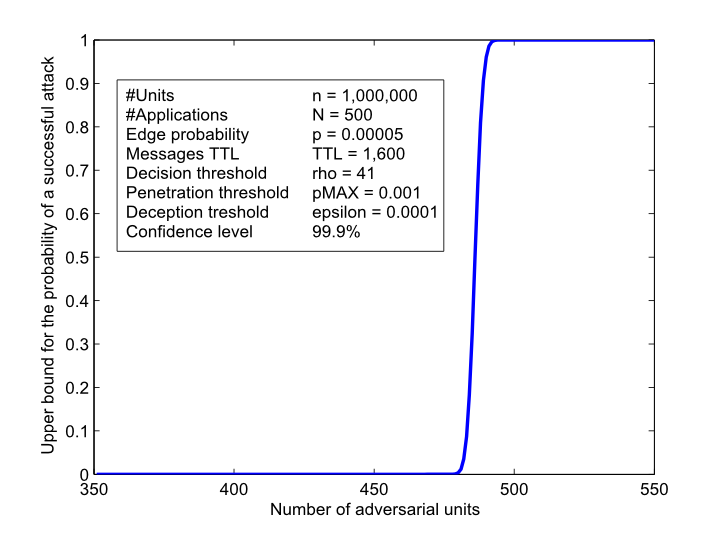}
\includegraphics[scale=0.35,bb=0 0 800 650]{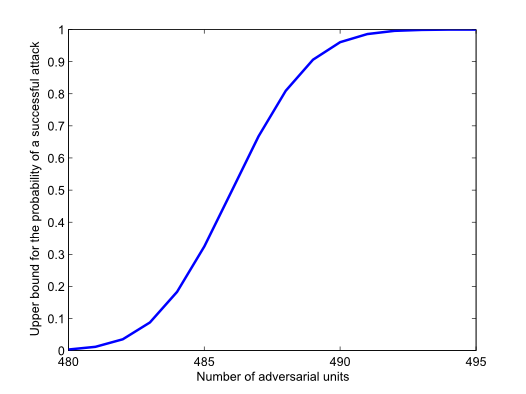}
\end{center}
\caption{An illustration of Theorem \ref{theorem.robust1} --- an upper bound for the success probability of a collaborative ``framing attack'' as a function of the number of adversarial devices. In this example, a changing number of adversaries are required to deceive at least 100 devices to think that some benign application is in fact malicious. Notice the phase transition point around 485 adversarial device.}
\label{fig.attack1new}
\end{figure}

\begin{figure}[htbp]
\begin{center}
\includegraphics[scale=0.5,bb=0 0 800 450]{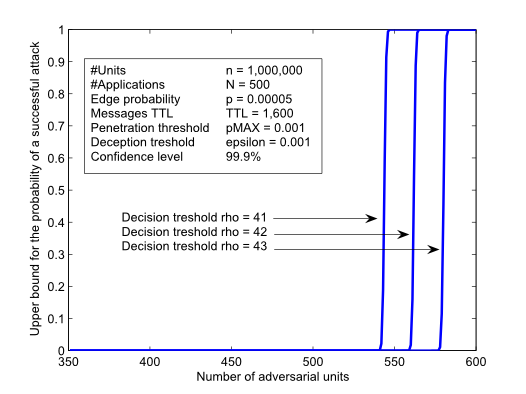}
\end{center}
\caption{An illustration of Theorem \ref{theorem.robust1} --- the influence of the value of decision threshold $\rho$ on the number of adversarial devices required for a successful attack. Note how as the adversaries increase their numbers, all that has to be done in order to prevent them from launching successful attacks is simply to slightly increase the value of the decision threshold $\rho$. This makes the \emph{TPP} algorithm economically favorable for networks operators. From experimental observations it can be seen that increasing the value of $\rho$ results only in slight increase in the vaccination time and network overhead, which converge asymptotically, whereas the resources required in order to launch stronger attacks grow approximately linearly. See more details in Figures~\ref{fig.attack1}}
\label{fig.attack3new}
\end{figure}

Theorem \ref{theorem.robust1} is illustrated in Figures \ref{fig.attack1new} and 
\ref{fig.attack3new}. Figure \ref{fig.attack1new} presents the execution of the vaccination algorithm in a network of 1,000,000 devices, where each device is connected to 50 neighbors. In this example we require that the number of devices that may be deceived by adversarial attacker would be at most 100. With the decision threshold $\rho$ properly calibrated according to Theorem \ref{theorem.robust1}, we show that as long as the number of adversaries is below 480 they cannot launch a successful attack. However, as the number of adversaries increases, such an attack quickly becomes a feasible option. Specifically, by increasing the number of adversaries by 3\% (from 480 to 494) the probability of a successful attack rises from 0 to 1.
In order to compensate the increase in adversaries number, devices operating the \emph{TPP} algorithm have to simply increase the value of $\rho$ by 1. Figure \ref{fig.attack3new} shows how each small increase in the value of the decision threshold $\rho$ requires the adversaries to increase their numbers by approximately 7\%. Interestingly, the effect of such changes in the value of $\rho$ on the algorithm's load and overhead is very small, as can be observed in Figure \ref{fig.attack1}.

Note that in the proof of Theorem \ref{theorem.robust1} we assumed that the adversarial devices may decide to send a false message concerning an application's maliciousness, but they must do so using the definitions of the \emph{TPP} algorithm. Namely, each device may send at most $p_{N} \cdot n$ messages, and send these messages to random destinations. If adversarial devices had been allowed flexibility in those constraints as well, a small group of adversarial devices could have sent an infinitely large number of false messages, that would have been propagated throughout the network, resulting in a successful attack (similarly to the case where $TTL \rightarrow \infty$).
Alternatively, had adversarial devices been allowed to chose the destination of the $p_{N} \cdot n$ messages they send, they could have all send them to the same destination $v$, thus guaranteeing that $v$ would be deceived. More generically, a group of $k = \i \cdot \rho$ of adversarial devices could have send their messages to $i \cdot p_{N} \cdot n$ different devices, guaranteeing their deception.

\begin{definition}
Let us denote by $P_{Attack-Destination} \left(TTL, \rho, \frac{k}{n}, \varepsilon\right)$ the attack success probability when adversarial devices are allowed to control the destination of the messages they produce.
\end{definition}

The following Corollary can be drawn~:

\begin{corollary}
\label{cor.robust1}
The probability that $k$ attackers that can control the destination of the $p_{N} \cdot n$ messages they produce will be able to make at least an $\epsilon$ portion of the network's devices treat some benign application $a_{i}$ as malicious, using the \emph{TPP} algorithm with a decision threshold $\rho$ is~:
\[
P_{Attack-Destination} \left(TTL, \rho, \frac{k}{n}, \varepsilon\right) \leq
P_{Attack} \left(TTL - 1, \rho, \frac{k}{n}, \varepsilon - \frac{k}{\rho} \cdot p_{N} \right)
\]
\end{corollary}

\section{Fault Tolerance to ``Leeching'' and ``Muting Attacks''}
\label{sec.leeching}

Collaborative systems, by their nature, are based on the fact that the participants are expected to allocate some of their resources (computational, energy, storage, etc') to be used by some collaborative algorithm. However, what happens when users decide to benefit from the system's advantages without providing the contribution that is expected from them ? This behavior, known as \emph{leeching} is a known artifact in many Peer-to-Peer data distribution systems, in which users often utilize the system for data download, without allocating enough upload bandwidth in return.
A similar behavior can also be the result of a deliberate attack on the system~---~a \emph{muting attack}. In this attack, one or more participants of the system block all the messages that are sent through them (namely~---~automatically decrease the TTL of those messages to zero). In addition, no original messages are being generate by these participants. The purpose of this attack is to compromise the correctness of the \emph{TPP} vaccination algorithm, which relies on the paths messages of a given TTL value are expected to perform.

Either performed as a way of evading the need to allocate resources for the collective use of the network, or maliciously as an adversarial abuse of the system, this behavior might have a significant negative influence on the performance of the \emph{TPP} algorithm. The operators of the network therefore need to have a way of calibrating the system so that it will overcome disturbances caused by any given number of participants who select to adopt this behavior.
In this section we show that in terms of completion time the \emph{TPP} algorithm is fault tolerant to the presence of blocking devices up to a certain limit.

\begin{definition}
Let $p_{mute}$ denote the probability that a given arbitrary network device stops generating vaccination messages or starts blocking some or all of the messages that are sent through it.
\end{definition}

We now show (Corollary \ref{cor.alg1.time-mute1}) that the expected vaccination time remains unchanged as long as~:
$p_{mute} \ll \frac{1}{\rho  \ln n}$.
For higher values of $p_{mute}$ Theorem \ref{theorem.timeout1-mute} presents an analytic upper bound for the algorithm's expected completion time.
For extremely high values of $p_{mute}$ the completion time of the algorithm is shown in Corollary \ref{cor.alg1.time-mute2} to be upper bounded by
$\frac{\frac{1}{4} p_{mute}}{1 - p_{mute}}  \cdot T_{Vac}^{2}$ (denoting by $T_{Vac}$ the vaccination time with the presence of no blocking devices).

As to the overall cost of the \emph{TPP} algorithm, it is shown in Corollary \ref{cor.alg1.cost-mute1} to remain completely unaffected by the presence of blocking devices, regardless of their number.
Due to space considerations, some of the proofs were omitted and can be found in the Appendix.

\begin{definition}
Let us denote by $T(n,p_{mute})$ the vaccination time of a network of $n$ devices, with a probability of $p_{mute}$ to block messages.
\end{definition}

\begin{theorem}
\label{theorem.timeout1-mute}
The vaccination completion time of the \emph{TPP} algorithm for some critical penetration $p_{MAX}$ in probability greater than $1 - \epsilon$, while at most $n \cdot p_{mute}$ devices may block messages forwarding and generation, is~:
\[T(n,p_{mute}) = \frac{2 (\rho -\ln \frac{\epsilon}{n} )}{1-e^{- \frac{1 - p_{mute} - e^{- timeout \cdot p_{mute}}}{p_{mute}} \cdot \frac{n \cdot p_{MAX} \cdot p_{N}}{2 T \cdot N \cdot (1 - E_{-}) ^{-1}}  }} \]
while for the calculation of $timeout$ we can use the expressions that appear in Theorem \ref{theorem.timeout1} or Theorem \ref{corthm.alg1.time}.
\begin{proof}
See Appendix for complete proof.
\end{proof}
\end{theorem}

We shall now observe the behavior of the expression above for various values of the ratio $r_{mute}$ defined as $p_{mute} \cdot timeout$. We shall note three complementary cases~: $r_{mute} \ll 1$ or $r_{mute} \gg 1$ or $r_{mute} \approx 1$.

It is easy to see that when $r_{mute} \ll 1$, the decay of the number of messages is negligible, namely~:
\[ \frac{1 - p_{mute} - e^{- timeout \cdot p_{mute}}}{p_{mute}} \approx timeout\]
As a result, Theorem \ref{theorem.timeout1-mute} can be approximated by Theorem \ref{theorem.timeout1}. Subsequently, the Theorems and Corollaries that are derived from Theorem \ref{theorem.timeout1} would hold, according to which we can see that~:
$p_{mute} \ll \frac{1}{O \left( \rho  \log n \right)}$.

Based on the above, we can now state the fault tolerance of the \emph{TPP} algorithm, with respect to the muting attack~:
\begin{corollary}
\label{cor.alg1.time-mute1}
The \emph{TPP} algorithm is fault tolerant with respect to the presence of $c \cdot O \left( \frac{n}{\rho  \ln n} \right)$ mute mobile devices (for some $c \ll 1$). Namely~:
\[T(n,p_{mute}) \approx T(n,0) \]
\end{corollary}

Let us now observe the case where $r_{mute} \gg 1$. In this case, most of the messages are likely to be blocked before completing their TTL-long path. This results in an increased vaccination time, as shown in the following Corollary~:
\begin{corollary}
\label{cor.alg1.time-mute2}
When the number of blocking devices is greater than $c \cdot O \left( \frac{n}{\rho  \ln n} \right)$ (for some $c \gg 1$), the completion time of the \emph{TPP} algorithm is affected as follows~:
\[T(n,p_{mute}) \leq \frac{\frac{1}{4} p_{mute}}{1 - p_{mute}}  \cdot T(n,0)^{2} \]
\begin{proof}
See Appendix for complete proof.
\end{proof}
\end{corollary}

Note that for every other value of $r_{mute}$ which revolves around 1, the expected vaccination time would move between $T(n,0)$ and $\frac{\frac{1}{4} p_{mute}}{1 - p_{mute}}  \cdot T(n,0)^{2}$ (see more details concerning the monotonic nature of $T(n,p_{mute})$ in the proof of Corollary \ref{cor.alg1.cost-mute1}).

Let us now examine the affect blocking devices may have on the number of messages sent throughout the execution of the algorithm. As shown in the following Corollary, the overall cost of the \emph{TPP} algorithm remains unaffected by the presence of any given number of blocking devices.

\begin{definition}
Let us denote by $M(n,p_{mute})$ the overall cost of the \emph{TPP} algorithm (messages sent + monitoring) for a network of $n$ devices, with a probability of $p_{mute}$ to block messages.
\end{definition}

\begin{corollary}
\label{cor.alg1.cost-mute1}
The overall cost of the \emph{TPP} algorithm is unaffected by the presence of blocking devices. Namely~:
\[ \forall p_{mute} < 1 \quad , \quad M(n,p_{mute}) = M(n,0)\]
\begin{proof}
See Appendix for complete proof.
\end{proof}
\end{corollary}

\section{Vaccination in General Graphs Using No Overlays}
\label{sec.broder}

An interesting question is raised when the forwarding of notification messages between the vertices is not assumed to be done using a random network overlay, but rather --- using only the edges of the network graph. This is also motivated by works such as \cite{Epidemic-Chakrabarti} and \cite{Garetto03modelingmalware}
which show how network topology may play an important role in the spreading of virus / malware (and subsequently, also of the vaccination messages).

In addition, we shall assume that we have no knowledge concerning the structure of the network graph, or its type. We now show that even in this case, the \emph{TPP} algorithm can still be used, albeit with a much larger completion time.

\begin{theorem}
\label{theorem.general.timeout}
Using the \emph{TPP} algorithm in an unknown graph with no network overlay, to guarantee a successful vaccination process for some critical penetration $p_{MAX}$, the following completion time is obtained~:

\[T_{Vac} = O \left(\left(\frac{T \cdot N}{p_{MAX} \cdot (1 - E_{-})}\right)^{\frac{2}{3}} \cdot n^{-\frac{2}{3}} \cdot \log^{\frac{4}{3}} n \right)\]
%
%
\begin{proof}
We shall first observe the following upper bound concerning the exploration time of a general graph using a decentralized group of $k$ random walkers \cite{exploration_broder}~:
\begin{displaymath}
E(ex_{G}) = O \left( \frac{|E|^{2} log^{3}n}{k^{2}} \right)
\end{displaymath}

Note that the coverage time of random walkers in graphs is also upper bounded by $\frac{4}{27} n^{3} + o(n^{3})$ \cite{Feige-random-upper1}. However, as we assume that $p_{N} < O(n^{-(0.5 + \epsilon)})$ (for some $\epsilon > 0$), using the bound of \cite{exploration_broder} gives a tighter bound.
Using our lower bound over the number of agents for $k$, and as $|E| = n \cdot p_{N}$ we get~:

\[
E(ex_{G}) = O \left( \frac{n^{2} p_{N}^{2} \log^{3}n}{timeout^{2} \cdot \frac{n^{4} \cdot p_{MAX}^{2} \cdot p_{N}^{2}}{T^{2} \cdot N^{2}} (1 - E_{-})^{2}} \right)
\]

Multiplying the exploration time by $\rho$ (in order to guarantee that each device will get at least $\rho$ messages) and replacing $E(ex_{G})$ with $timeout$ we see that~:

\[T_{Vac} = O \left(\sqrt[3]{\frac{\rho \cdot \left(\frac{T \cdot N}{p_{MAX} \cdot (1 - E_{-})}\right)^{2}}{n^{2}}}  \cdot \log(n) \right)\]

Assuming also that $\rho = O(\ln n)$, and the rest is implied.
\end{proof}
\end{theorem}

\noindent The overall cost of the algorithm in graphs with no overlays is given in Theorem \ref{theorem.general.cost}.
\begin{theorem}
\label{theorem.general.cost}
Using the \emph{TPP} algorithm in an unknown graph with no network overlay, to guarantee a successful vaccination for some penetration threshold $p_{MAX}$, the overall cost of the algorithm is upper bounded by the following expression~:
\[
O\left(\left(\frac{T \cdot N}{p_{MAX} \cdot (1 - E_{-})}\right)^{\frac{1}{3}} \cdot n^{\frac{2}{3}} \cdot \log^{\frac{8}{3}} n
 \cdot C_{S} +
\frac{n^{\frac{7}{3}} \cdot \log^{\frac{4}{3}} n }{|E| \cdot \left(\frac{T \cdot N}{p_{MAX} \cdot (1 - E_{-})}\right)^{\frac{1}{3}} }
 \cdot C_{M}\right)
\]
%
%
\begin{proof}
As to the overall cost of the algorithm, we know that~:
\[
M = O\left( k \cdot timeout \cdot C_{S} + \frac{k}{X} \cdot C_{M}\right)
\]
Using the expressions for $timeout$, $k$ and $p_{N}$, the rest is implied.
%
%
\end{proof}
\end{theorem}

\section{Experimental Results}
\label{sec.experimental}

We have implemented the \emph{TPP} algorithm in a simulated environment and tested its performance in various scenarios. Due to space considerations, we present in this Section only a sample of the results we have obtained. We have tested a network of $n = 1000$ devices, having access to $N = 100$ applications, one of which was malicious\footnote{Note that the number of malicious applications does not influence the completion time of algorithm, as monitoring and notification is done in parallel. The number of message, however, grows linearly with the number of malicious applications.}. We assume that each device downloads 30 random applications, monitors 1 application every week, and can send notification messages to 10 random network members (namely, $p_{N} = 0.01$). We require that upon completion, at least 990 network members must be aware of the malicious application (namely, $p_{MAX} = 0.01$), with $\epsilon = 0.001$.
In addition, we assumed that among the network members there are 100 adversaries, trying to deceive at least 50 mobile devices to believe that some benign application is malicious.

\begin{figure}[htbp]
\begin{center}
\includegraphics[scale=0.3,bb=0 0 800 450]{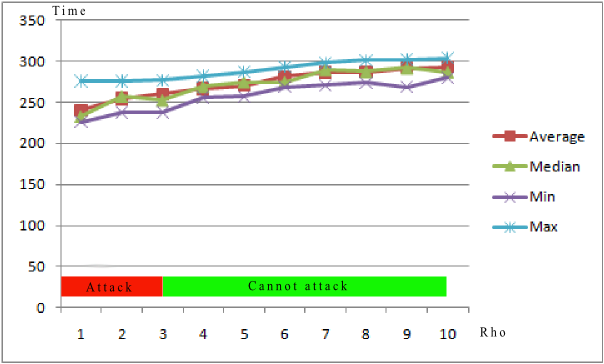}
\includegraphics[scale=0.3,bb=0 0 800 450]{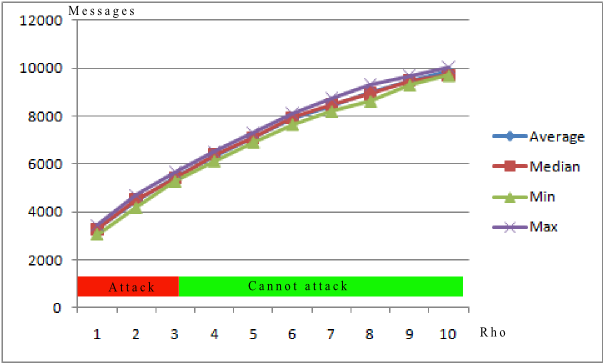}
\end{center}
\caption{An experimental result of a network of $n = 1000$ members, with $N = 100$ applications, $p_{MAX} = 0.01$, $p_{N} = 0.01$ and 100 adversaries that try to mislead at least $5\%$ of the network into believing that some benign application is malicious. Notice how changes in $\rho$ dramatically effect the adversaries' success probability, with almost no effect on the completion time.}
\label{fig.attack1}
\end{figure}

Figure~\ref{fig.attack1} shows the time (in days) and messages required in order to complete this monitoring, as a function of the decision threshold $\rho$. We can see that whereas the adversaries succeed in probability 1 for $\rho < 3$, they fail in probability 1 for any $\rho \geq 3$. Note the extremely efficient performance of the algorithm, with completion time of $\sim 260$ days using only 5 messages and at most 30 monitored applications per user. The same scenario would have resulted in 100 messages per user using the conventional \emph{flooding} algorithm, or alternatively, in 700 days and 100 monitored applications per user using a non-collaborative scheme.
Notice how the increase in the time and network overhead as a result of an increase in $\rho$ (in order to compensate an increase in the number of adversaries) converge asymptotically. This makes the algorithm economically scalable, as discussed in Section~\ref{sec.robust}.

\begin{figure}[htbp]
\begin{center}
\includegraphics[scale=0.3,bb=0 0 800 450]{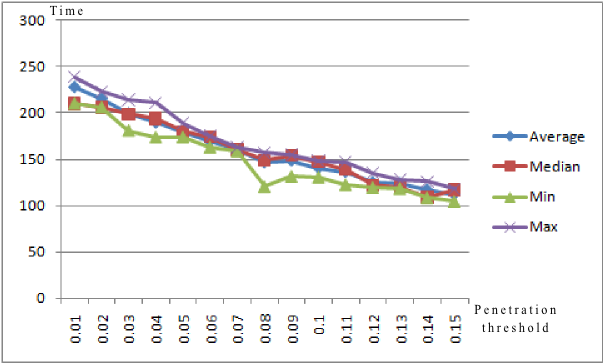}
\includegraphics[scale=0.3,bb=0 0 800 450]{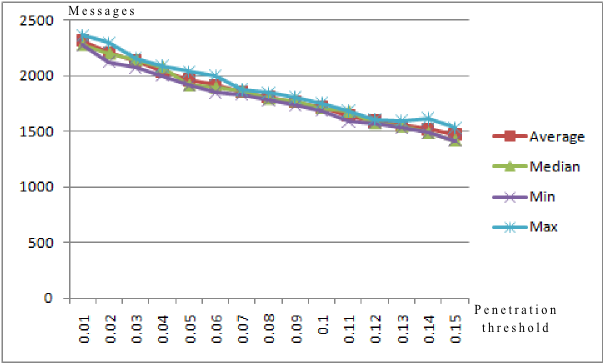}
\end{center}
\caption{The effect of decreasing the penetration threshold $p_{MAX}$ on the algorithm's completion time and number of messages ($\rho = 1$).}
\label{fig.attack2}
\end{figure}

\begin{figure}[htbp]
\begin{center}
\includegraphics[scale=0.35,bb=0 0 600 450]{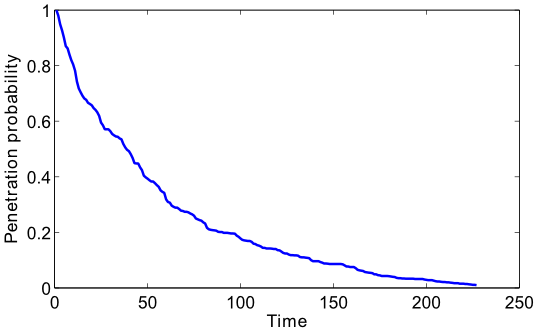}
\includegraphics[scale=0.35,bb=0 0 600 450]{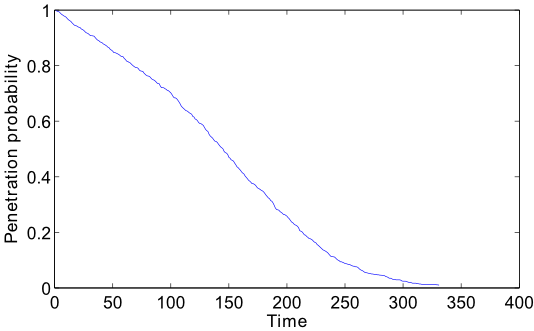}
\end{center}
\caption{The \emph{penetration probability} of the malicious application, as a function of the time, with $\rho = 1$ (on the left) and $\rho = 20$ (on the right).}
\label{fig.attack4}
\end{figure}

\begin{figure}[htbp]
\begin{center}
\includegraphics[scale=0.4,bb=0 0 600 450]{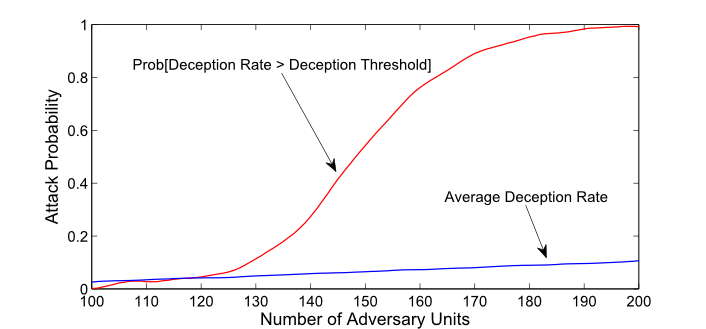}
\end{center}
\caption{An illustration of the phase transition that is displayed when observing the influence of the number of adversaries over their probability to successfully generate an attack.
Note how an increase of $20\%$ in the number of adversaries increases the probability to deceive a large enough portion of the network from less than 0.2 to approximately 0.8.}
\label{fig.attack-transition1}
\end{figure}

Figure \ref{fig.attack2} demonstrates the decrease in completion time and network overhead as a result of increasing the penetration threshold $p_{MAX}$.
Figure \ref{fig.attack4} demonstrates the evolution in the malicious application's penetration probability throughout the vaccination process. An interesting phenomenon is demonstrated in Figure~\ref{fig.attack-transition1}, where the number of adversarial devices is gradually increased, and their success in deceiving $5\%$ of the network's members is studied. As the deception rate increases linearly, the success to generate a successful attack displays a phase transition~---~growing rapidly from ``a very low attack success probability'' to ``very high attack success probability'' with the increase of only $20\%$ in the number of adversaries.

\setlength{\unitlength}{1cm}

\setlength{\unitlength}{1cm}

\begin{figure}[htbp]
\begin{center}
\textbf{With No Collaborative Monitoring}
\qquad \qquad
\textbf{With \emph{TPP} Vaccination Mechanism}
\includegraphics[scale=0.3,bb=0 0 800 400]{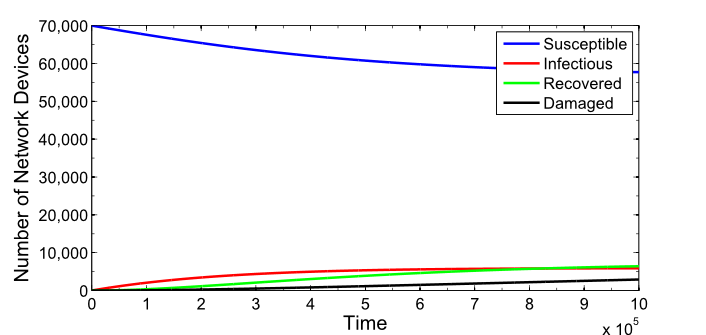}
\includegraphics[scale=0.3,bb=0 0 800 400]{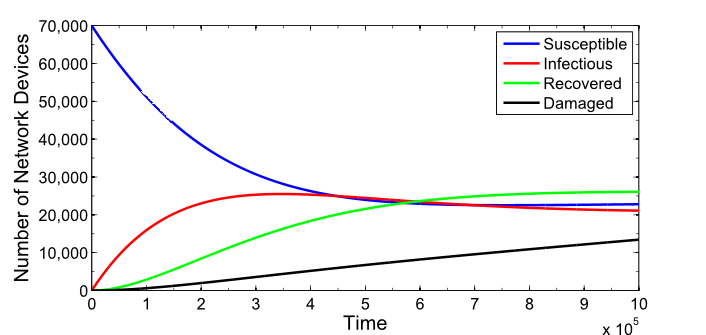}
\end{center}
\caption{An experimental demonstration of the benefits of using the proposed collaborative monitoring algorithm. Implemented on a network that is based on real-life mobile network of 70,000 devices, the \emph{TPP} algorithm significantly decreases both the number of infectious devices as well as the accumulated number of malicious incidents at the network (i.e. damages caused by activations of malicious applications that were installed on mobile devices).}
\label{fig.simulationepidemic1}
\end{figure}

As stated in previous sections, the main contribution that stems from implementing the \emph{TPP} algorithm is the fact that the vast majority of the mobile devices become immune to any new malicious applications that may be introduced to the network. We have implemented a network of 70,000 units based on the aggregated calls-graph that was received from an actual mobile network. Our simulation assumed that each user downloads an average of 50 applications per month, out of which $\eta_{n} = 5$ new applications. To this model we have periodically injected a small number of malicious applications (approximately 0.5\% of the applications). In this simulation we used the well known \emph{SIR} epidemic model, with a slight variation --- as the network is dynamic, we assumed that there is a constant flow of users leaving the network, which is compensated by a flow of users joining the network. This allows the network to avoid the ``everyone dies eventually'' syndrome, by constantly trading a portion of the recovered devices for new susceptible ones. Notice that a device may become infected either as a result of downloading a malicious application or as a result to an exposure to an infected device. In both cases, a mandatory condition for the infection is that the device was not vaccination against this application already.
Figure~\ref{fig.simulationepidemic1} presents the dynamics of the network, comparing the number of infected and damaged devices when the collaborative vaccination mechanism is present to the number of infected and damaged mobile devices without it. Notice how the proposed algorithm decreases the number of infectious devices at the steady state of the network, as well as the accumulated number of malicious incidents, by approximately 75\%.

\section{Conclusions and Future Work}
\label{sec.conc}

In this work we have presented the \emph{TPP} TTL-based propagation algorithm, capable of guaranteeing the collaborative vaccination of mobile network users against malicious applications. The performance of the algorithm was analyzed and shown to be superior compared to state of the art in this field, guaranteeing higher number of suspicious applications that can be monitored concurrently, and do so using a lower network overhead.

The algorithm was also shown to be capable of overcoming the presence of adversarial devices who try to inject messages of false information into the network. The challenge of filtering out false information which is injected into a collaborative information propagation networks resembles the known ``faulty processors'' problem. For example, the following work discusses the challenge of synchronizing the clock of a communication network of size $n$ when $\frac{n}{3}$ faulty processors are present \cite{Faulty-Halevi}. Another interesting work in this scope is the work of \cite{Tsudik1} which discusses a collaborative fault tolerant ``acknowledgement propagation'' algorithm, for reporting on the receipt (or lack of) of sent messages.

It should be noted that during the analysis of the fault tolerance of the algorithm, we assumed that although an attacker can send reports of benign applications, or alternatively --- refuse to forward messages passed through it, the basic parameters of the algorithm are still preserved. Namely, adversarial devices are not allowed to send more than $X$ messages or generate messages with $TTL$ values higher than the value allowed by the network operator. In addition, we assumed that while an adversary can ``kill'' messages sent to it by refusing to forward them, it cannot however interfere with the content of the message itself (e.g. by changing the identity of the message's original sender, its current $TTL$, or the identity of the application it reports of). The exact implementation details of a cryptographic protocol of these properties, however, is out of the scope of this work,

Future versions of the work should investigate the generalization of the propagation mechanism, allowing the number of alert messages generated to be dynamically calibrated in order to further decrease the algorithm's overhead (for example, as a function of the number of alerts already received concerning the application). Alternatively, upon forwarding a message, devices might be requested to send more than a single copy of the message they received.

Another interesting topic to investigate is the use of the mechanism proposed in this work as an infrastructure for other security related problems. One example can be the problem of collaboratively coping with malicious beacons in hostile wireless environment, as discussed in \cite{Voting-Liu}.
Most of the existing localization protocols for sensor networks are vulnerable in hostile environments, requiring for the enhancement of the security of location discovery. The work of \cite{Voting-Liu} presents voting-based methods to tolerate malicious attacks against range-based location discovery in sensor networks. This problem seems to benefit from the use a mechanism which is fault tolerant to the injection of false information, such as the algorithm we propose in this work.

\appendix

\section{Proof of Theorem \ref{theorem.m}}

\textbf{THEOREM \ref{theorem.m}}
The time it takes $k$ random walkers to complete a $\rho$-coverage of $G$ in probability greater than $1-\epsilon$ (denoted as T(n)) can be bounded as follows~:
\begin{displaymath}
\frac{2 \left(\rho -\ln \frac{\epsilon}{n} \right)}{1-e^{-\frac{3 k}{2 n (1 - \frac{1}{\ln n})}}} \leq T(n) \leq \frac{2 \left(\rho -\ln \frac{\epsilon}{n} \right)}{1-e^{-\frac{k}{2n}}}
\end{displaymath}
\begin{proof}
Since our bounds are probabilistic, we can state that the following ``bad events'' occur with very low probability (e.g.
$2^{-\omega(n)}$).
Event $E_{low \  degree}$, defined as the existence of a
vertex $v \in V$ with $deg(v) < \frac{n \cdot p_{N}}{2}$.
Event $E_{high \  degree}$, defined as the existence of a
vertex $v \in V$ with $deg(v) > \frac{3 \cdot n \cdot p_{N}}{2}$.
Using the \emph{Chernoff} bound on $G$ we get~:
\[prob[deg(v) < \frac{n \cdot p_{N}}{2}] < e ^{-\frac{n \cdot p_{N}}{8}}\]
Applying union bound on all vertices we get~:
\[
    Prob[E_{low \ degree}] <
    n \cdot e ^{-\frac{n \cdot p_{N}}{8}} < 2^{-\omega(n)}
\]
Similarly~:
%
%
\[
    Prob[E_{high \ degree}] <
     2^{-\omega(n)}
\]

From now on we assume that $E_{low \ degree}$ and $E_{high \ degree}$ do not occur, and condition all probabilities over this assumption.
In the private case of $\forall v \in V \ , \ deg(v) = p_{N} \cdot n$, every analysis that is based on the expected number of neighbors shall hold.

\remove{
In order to continue analyzing the execution process of the \emph{TPP} algorithm we show that the locations of the notification messages at any given time are purely random.
\begin{lemma} \label{lemma_rand_placement}
The placement of the agents after every-step is purely random over the nodes
of $G$.
\begin{proof}
This is true since
\begin{enumerate}
\item  The initial placement of the agents is random
\item  Their movement is random
\item  The graph $G$ is random
\end{enumerate}
\end{proof}
\end{lemma}
} 
In order to continue analyzing the execution process of the \emph{TPP} algorithm we note that as
the initial placement of the agents is random, their movement is random and the graph $G$ is random, we can see that the placement of the agents after every-step is purely random over the nodes.
Using these observation, the number of agents residing in adjacent vertices from some vertex $v$ can be produced~:

\begin{lemma}
Let $v \in V$ be an arbitrary vertex of $G$. Let $N_1(v,t)$ be the number of
agents which reside on one of $Neighbor(v)$ (adjacent vertices to $v$) after step $t$ :
\[\forall t \geq 0 : \frac{3 \cdot p_{N} \cdot k}{2} \geq E[N_1(v,t)] \geq \frac{p_{N} \cdot k}{2}\]

In other words, the expected number of agents who reside in distance $1$ from $v$ after every step
is at least $\frac{p_{N} \cdot k}{2}$ and at most $\frac{3 \cdot p_{N} \cdot k}{2}$.
\begin{proof}
Upon our assumption, in $G(n,p_{N})$ the number of incoming neighbors for some vertex $v$ is
at least $\frac{1}{2} p_{N} \cdot n$ and at most $\frac{3}{2} p_{N} \cdot n$. In addition, 
for every $u \in V(G)$, $Prob[$some agent resides on $u] = \frac{k}{n}$. Combining the two we see that~:
\[\forall t \geq 0 : E[N_1(v,t)] \geq \frac{p_{N} \cdot n \cdot k}{2n} \geq \frac{1}{2} p_{N} \cdot k\]
\[\forall t \geq 0 : E[N_1(v,t)] \leq \frac{3p_{N} \cdot n \cdot k}{2n} \geq \frac{3}{2} p_{N} \cdot k\]
\end{proof}
\end{lemma}

\begin{lemma}
\label{lemma.prob1}
For any vertex $v \in V$, the probability of $v$ being notified at
the next time-step that $a_{i}$ is malicious is at least $1-e^{-\frac{k}{2n}}$ and at most $1 - e^{-\frac{3 k}{2 n (1 - \frac{1}{\ln n})}}$.
\begin{proof}
The probability that an agent located on a vertex $u$ such that $(u,v) \in E$ will move to $v$ at the next time-step is $\frac{1}{p_{N} \cdot n}$. The number of agents that are located in adjacent vertices to $v$ is between $\frac{k}{2} p_{N}$ and $\frac{3k}{2} p_{N}$. Therefore, the probability that $v$ will not be reported about $a_{i}$ at the next time-step is between $(1-\frac{1}{p_{N} \cdot n})^{\frac{1}{2} p_{N} \cdot k}$ and
$(1-\frac{1}{p_{N} \cdot n})^{\frac{3}{2} p_{N} \cdot k}$. Using the well known inequality $(1-x) < e^{-x}$ for
$x < 1$, we can bound this probability from above
by~:
\[(e^{-\frac{1}{p_{N} \cdot n}})^{\frac{1}{2} p_{N} \cdot k} \leq e^{-\frac{k}{2n}}\]
Using the inequality $(1-x) > e^{-\frac{x}{1-x}}$ for $x < 1$, it will be bounded from below by~:
\[\left(e^{-(p_{N} \cdot n)^{-1} (1-\frac{1}{p_{N} \cdot n})^{-1}}\right)^{\frac{3}{2} p_{N} \cdot k} \geq e^{-\frac{3 p_{N} k}{2 p_{N} (n - \frac{1}{p_{N}})}} \geq e^{-\frac{3 k}{2 n (1 - \frac{1}{\ln n})}}  \]
In order to guarantee that the graph is connected we know that $P_{N} \geq \frac{\ln n}{n}$ \cite{ER_Graphs2}.
Therefore, the probability that $v$ will be notified on the next time-step is at least $1-e^{-\frac{k}{2n}}$ and at most $1 - e^{-\frac{3 k}{2 n (1 - \frac{1}{\ln n})}}$.
\end{proof}
\end{lemma}
Interestingly, this fact holds for any positive $p_{N}$ (the density parameter of $G$).

Lemma~\ref{lemma.prob1} states the probability that some vertex $v \in V$ will be reported of $a_{i}$ at
the next time-step. This is in fact a Bernoulli trial with probability $p_{success}$.
Now we bound the probability of failing this trial (not notifying vertex $v$ enough times)
after $m$ steps. Let $X_v(m)$ denote the number of times that a notification message had arrived to $v$
after $m$ steps, and let $F_v(m)$ denote the event that $v$ was not notified enough times
after $m$ steps (i.e. $X_v(m) < \rho$). We additionally denote by $F(m)$ the
event that one of the vertices of $G$ where not notified enough times after $m$ steps (i.e.
$\bigcup_{v \in V(G)} F_v(m)$). We use the \emph{Chernoff} bound~:
\[ P[X_v(m) < (1-\delta)p_{success}m] < e^{-\delta^2 \frac{m p_{success}} {2}}\]
in which we set $\delta = 1-\frac{\rho}{mp_{success}}$. We can then see that~:
\[ P[X_v(m) < \rho]
 < e^{-(1-\frac{\rho}{m p_{success}})^2 \frac{m p_{success}}{2}}\]
namely~:
$P[F_v(m)] <  e^{\rho - \frac{m p_{success}}{2}}$.
Applying the union bound we get~:
\[P[e_1 \cup e_2 \cup \ldots \cup e_n] \leq P[e_1] + P[e_2] + \ldots + P[e_n]\]
on all $n$ vertices of $G$. Therefore we can bound the probability of failure on any vertex $v$ (using Lemma~\ref{lemma.prob1}) as follows~:
\[Pr[F(m)] \leq n e^{\rho -\frac{m p_{success}}{2}} \leq n e^{\rho -\frac{m (1-e^{-\frac{k}{2n}})}{2}} \leq \epsilon \]

Assigning $p_{success}=1-e^{-\frac{k}{2n}}$ (worse case) and $p_{success}= 1 - e^{-\frac{3 k}{2 n (1 - \frac{1}{\ln n})}}$ (best case),
the rest is implied.
\end{proof}

\section{Leeching and Muting Attacks}

In this section we give the complete proofs for the Theorems and Corollaries that appear in Section \ref{sec.leeching} of the main body of the paper.

\noindent \textbf{Detailed Analysis --- Completion Time}.
Recall we denote by $p_{mute}$ the probability that a given network device may decide to stop generating vaccination messages and block some or all of the messages that are received by it. In addition, recall that $T(n,p_{mute})$ denotes the vaccination time of a network of $n$ devices, with a probability of $p_{mute}$ to block messages. Theorem \ref{theorem.timeout1} can now be revised in the following way~:

\textbf{THEOREM \ref{theorem.timeout1-mute} : }
The vaccination completion time of the \emph{TPP} algorithm for some critical penetration $p_{MAX}$ in probability greater than $1 - \epsilon$, while at most $n \cdot p_{mute}$ devices may block messages forwarding and generation, is~:
\[T(n,p_{mute}) = \frac{2 (\rho -\ln \frac{\epsilon}{n} )}{1-e^{- \frac{1 - p_{mute} - e^{- timeout \cdot p_{mute}}}{p_{mute}} \cdot \frac{n \cdot p_{MAX} \cdot p_{N}}{2 T \cdot N \cdot (1 - E_{-}) ^{-1}}  }} \]
while for the calculation of $timeout$ we can use the expressions that appear in Theorem \ref{theorem.timeout1} or Theorem \ref{corthm.alg1.time}.
\begin{proof}
The expected number of new messages created at time $t$, $\hat{k}(t)$ is expected to be~:
\[\hat{k}(t) = (1 - p_{mute}) \frac{n^{2} \cdot p_{a_{i}} \cdot p_{N}}{T \cdot N} (1 - E_{-})\]

Given a message created with a TTL of $timeout$, at every time step it has a probability of $p_{mute}$ to be sent to a network node which will not forward it onwards. Therefore, given a group of $m$ messages, created at time $t$. Then at time $t+i$ (for every $i < timeout$), only $(1 - p_{mute})^{i} \cdot m$ messages would remain alive. The average number of messages at any time step between $t$ and $t + timeout$ is therefore~:
\[
m \frac{1 - (1 - p_{mute})^{timeout}}{timeout \cdot p_{mute}}
\]

As we assume that $\forall t < T_{Vaccination} \ p_{a_{i}} \geq p_{MAX}$, the number of agents $k$ would be at least~:
\[\frac{1 - p_{mute} - (1 - p_{mute})^{timeout + 1}}{p_{mute}} \cdot \frac{n^{2} \cdot p_{MAX} \cdot p_{N}}{T \cdot N} (1 - E_{-})\]

Using again the fact that $\forall x < 1 \ (1-x) < e^{-x}$, we can see that~:
\[k > \frac{1 - p_{mute} - e^{- timeout \cdot p_{mute}}}{p_{mute}} \cdot \frac{n^{2} \cdot p_{MAX} \cdot p_{N}}{T \cdot N} (1 - E_{-})\]

Recalling Theorem \ref{theorem.m}~:
\[T(n,0) = \frac{2 (\rho -\ln \frac{\epsilon}{n} )}{1-e^{-\frac{k}{2n}}}\]
and assigning the revised expression for $k$, the rest is implied.
\end{proof}

\textbf{COROLLARY \ref{cor.alg1.time-mute2} : }
When the number of blocking devices is greater than $c \cdot O \left( \frac{n}{\rho  \ln n} \right)$ (for some $c \gg 1$), the completion time of the \emph{TPP} algorithm is affected as follows~:
\[T(n,p_{mute}) \leq \frac{\frac{1}{4} p_{mute}}{1 - p_{mute}}  \cdot T(n,0)^{2} \]
\begin{proof}
When $r_{mute} \gg 1$ Theorem \ref{theorem.timeout1-mute} converges as follows~:
\begin{observation}
\label{obs.alg1.time-mute2}
When $p_{mute} \gg timeout^{-1}$, the vaccination time of the \emph{TPP} algorithm is~:
\[T(n,p_{mute}) = \frac{2 (\rho -\ln \frac{\epsilon}{n} )}{1-e^{- \frac{1 - p_{mute}}{p_{mute}} \cdot \frac{n \cdot p_{MAX} \cdot p_{N}}{2 T \cdot N} (1 - E_{-})   }} \]
\end{observation}

Assuming again that
$\epsilon = n^{-\alpha} \quad s.t. \quad \alpha \in \mathbb{Z}^{+}$
then provided that~:
\[\frac{1 - p_{mute}}{p_{mute}} \cdot \frac{n \cdot p_{MAX} \cdot p_{N}}{2 T \cdot N} (1 - E_{-}) < 1\]
we see that~:
\[T(n,p_{mute}) = \frac{4 N \cdot T \cdot p_{mute} (\rho + (\alpha + 1) \ln n )}{(1 - p_{mute}) \cdot n \cdot p_{MAX} \cdot p_{N} (1 - E_{-})} \]

Using Theorem \ref{corthm.alg1.time}, we shall calculate $\Delta_{p_{mute}}$, denoting the increase in the vaccination time as a result of the presence of the blocking nodes~:

\[
\Delta_{p_{mute}} = \frac{T(n,p_{mute})}{T(n,0)} = \frac
{\frac{4 N \cdot T \cdot p_{mute} (\rho + (\alpha + 1) \ln n )}{(1 - p_{mute}) \cdot n \cdot p_{MAX} \cdot p_{N} (1 - E_{-})}}
{4 \sqrt{\frac{T \cdot N \left(\rho + (\alpha + 1) \ln n \right)}{n \cdot p_{MAX} \cdot p_{N} \cdot (1 - E_{-})}}}
\]
and after some arithmetics we can see that~:
\[
\Delta_{p_{mute}} = \frac{T(n,p_{mute})}{T(n,0)} = \frac{\frac{1}{4} p_{mute}}{1 - p_{mute}}  \cdot T(n,0)
\]
\end{proof}

\noindent \textbf{Detailed Analysis --- Cost}.
Recall we denote by $M(n,p_{mute})$ the overall cost of the \emph{TPP} algorithm (messages sent + monitoring) for a network of $n$ devices, with a probability of $p_{mute}$ to block messages. As shown in the following Corollary, the overall cost of the \emph{TPP} algorithm remains unaffected by the presence of any given number of blocking devices.

\textbf{COROLLARY \ref{cor.alg1.cost-mute1} : }
The overall cost of the \emph{TPP} algorithm is unaffected by the presence of blocking devices. Namely~:
\[ \forall p_{mute} < 1 \quad , \quad M(n,p_{mute}) = M(n,0)\]
\begin{proof}
We have already shown before (Corollary \ref{cor.alg1.time-mute1}) that when the number of blocking devices is smaller than $c \cdot O ( \frac{n}{\rho  \ln n} )$ (for some $c \ll 1$), the system is approximately unaffected by the blocking devices. Therefore, we can expect that~:
\[M(n,p_{mute}) \approx M(n,0) \]

Interestingly, this is also the case when the number of blocking devices is far greater.
Recalling Observation~\ref{obs.simple.timeout.results} the cost of the algorithm with no blocking devices equals~:
\[ M(n,0) = O \left(k \cdot T(n,0) + \frac{k}{n \cdot p_{N}} \cdot C_{M} \right) \]

Denoting by $k_{(n,p_{mute})}$ the expected number of active agents at each time step, we have already shown in the proof of Theorem \ref{theorem.timeout1-mute} that for large values of $p_{mute}$~:
\[
k_{(n,p_{mute})} \approx \frac{1- p_{mute}}{p_{mute}} \cdot T(n,0)^{-1} \cdot k_{(n,0)}
\]

The overall cost of the algorithm assuming large number of blocking devices equals~:
\[ M(n,p_{mute}) = O \left(k_{(n,p_{mute})} \cdot T(n,p_{mute}) + \frac{k_{(n,p_{mute})}}{n \cdot p_{N}} C_{M} \right)\]

Assigning the values of $k_{(n,p_{mute})}$ and $T(n,p_{mute})$ we can now see that~:
\[ M(n,p_{mute}) = O \left(k_{(n,0)} \cdot T(n,0) + \frac{   \frac{1- p_{mute}}{p_{mute}} \cdot k_{(n,0)}    }{T(n,0) \cdot n \cdot p_{N}} C_{M} \right)\]

As we assumed that $p_{mute} \gg \frac{1}{\rho \ln n}$
we know that~:
\[\frac{1- p_{mute}}{p_{mute}}  \ll   \frac{1- \frac{1}{\rho \ln n}}{\frac{1}{\rho \ln n}}  \ll  \rho \ln n - 1 \]

The expression for the overall cost of the algorithm can now be rewritten as~:
\[ M(n,p_{mute}) = O \left(k_{(n,0)} \cdot T(n,0) + \frac{   \rho \ln n \cdot k_{(n,0)}    }{T(n,0) \cdot n \cdot p_{N}} C_{M} \right)\]
and as $T(n,0) = \Omega(\ln n)$, we get the requested result of~:
\[ M(n,p_{mute}) = O \left(k_{(n,0)} \cdot T(n,0) + \frac{   k_{(n,0)}    }{n \cdot p_{N}} C_{M} \right) = M(n,0)\]

We have shown that $M(n,p_{mute}) = M(n,0)$ for $p_{mute} \ll timeout^{-1}$ and well as for $p_{mute} \gg timeout^{-1}$. We shall now upper bound the overall cost of the algorithm for values of $p_{mute}$ which are close to $timeout^{-1}$. For this, we shall find the value of $p_{mute}$ which maximizes $M(n,p_{mute})$~:

\[
\frac{\partial M(n,p)}{\partial p} = \frac{\partial k_{(n,p)}}{\partial p}  T(n,p) + \frac{\partial T(n,p)}{\partial p}  k_{(n,p)} + \frac{\partial k_{(n,p)}}{\partial p} \frac{C_{M}}{n \cdot p_{N}}
\]

Recalling (from the proof of Theorem \ref{theorem.timeout1-mute}) that~:
\[k > \frac{1 - p_{mute} - e^{- timeout \cdot p_{mute}}}{p_{mute}} \cdot \frac{n^{2} \cdot p_{MAX} \cdot p_{N}}{T \cdot N} (1 - E_{-})\]
we see that~:
\[
\frac{\partial k_{(n,p)}}{\partial p} = \frac{(p \cdot timeout + 1) \cdot e^{- timeout \cdot p}  - 1  }{p^{2}} \cdot \alpha
\]
where:~
\[\alpha = \frac{n^{2} \cdot p_{MAX} \cdot p_{N}}{T \cdot N} (1 - E_{-})\]

Therefore~:
\[
\forall p > 0 \quad , \quad \frac{\partial k_{(n,p)}}{\partial p} < 0
\]
and the number of agents is monotonously decreasing.

Examining the behavior of the cleaning time, we can see that~:
\[\frac{\partial T(n,p)}{\partial p} =
- \frac{\partial k_{(n,p)}}{\partial p} \cdot \frac{2 e^{- \frac{1 - p - e^{- timeout \cdot p}}{p} \cdot \beta}}
{n (1-e^{- \frac{1 - p - e^{- timeout \cdot p}}{p} \cdot \beta  })^{2}}
\]
where~:
\[\beta = \frac{n \cdot p_{MAX} \cdot p_{N}}{2 T \cdot N \cdot (1 - E_{-}) ^{-1}}\]

Using our observation concerning $\frac{\partial k_{(n,p)}}{\partial p}$, we see that~:
\[
\forall p > 0 \quad , \quad \frac{\partial T(n,p)}{\partial p} > 0
\]
and the cleaning time is monotonously increasing.

Returning to the derivative of the overall price function, let us divide it to two components. The first component representing the number of messages sent during the algorithm while the second representing the monitoring activities of the devices~:
\[
\frac{\partial M(n,p)}{\partial p} = M_{1}(p) + M_{2}(p)
\]
where~:
\[M_{1}(p) = \frac{\partial k_{(n,p)}}{\partial p}  T(n,p) + \frac{\partial T(n,p)}{\partial p}  k_{(n,p)}\]
and~:
\[M_{2}(p) = \frac{\partial k_{(n,p)}}{\partial p} \frac{1}{n \cdot p_{N}} \cdot C_{M}\]

As $k_{(n,p)}$ is monotonously decreasing, the cost of the algorithm due to monitoring activities, represented by $M_{2}(p)$, would be maximized when $p_{mute} = 0$ (for which we have already shown that $M(n,p) = M(n,0)$).

As to $M_{1}(p)$, it can be written as~:
\[M_{1}(p) = \frac{\partial k_{(n,p)}}{\partial p}  \left( T(n,p) -
\frac{e^{- \beta \cdot x_{p}}}
{(1-e^{- \beta \cdot x_{p} })^{2}}
\cdot x_{p} \cdot \beta \right)
\]
where~:
\[\beta = \frac{n \cdot p_{MAX} \cdot p_{N}}{2 T \cdot N \cdot (1 - E_{-}) ^{-1}}\]
and~:
\[x_{p} = \frac{1 - p - e^{- timeout \cdot p}}{p}\]

Studying this function reveals that it has a single minimum point, while $M_{1}(0) = 0$ and $M_{1}(1) \rightarrow \infty$. This means that the maximal values of $M(n,p_{mute})$ are received either at $p_{mute} = 0$ or at $p_{mute} = 1$. As we have already shown that at these points the overall cost of the algorithm remains unchanged, we can conclude that the overall cost of the algorithm at any point between these values of $p_{mute}$ is also bounded by $M(n,0)$.
\end{proof}

\section{$\lambda$ Optimization}
%
Recall that while analyzing the performance of the \emph{TPP} algorithm we artificially divided the vaccination process into two phases. At the first phase, the vaccinating agents are generated, while the proliferation of the agents is done in the second phase. The ratio between the two phases was assumed to be 1, namely~---~we assumed that the two phases are identical in length. Following is the analysis of the selection of this ratio, which demonstrates that this is indeed the optimal division.

Let us recall that~:
\[
    timeout = \lambda \cdot (T_{Generation} + timeout)
\]
and that~:
\[
    T_{Generation} = \frac{1 - \lambda}{\lambda} \cdot timeout
\]

As we assumed in the previous section that $\lambda = 0.5$, it is interesting to know whether another value of $\lambda$ (perhaps, a value that depends of the properties of the network) may yield a lower completion time.

Concerning $T_{Vac}$, it can now be seen that~:
\[T_{Vac} \leq \frac{1}{\lambda} \cdot timeout \]

%
%
%
Similar to the calculation of the number of agents $k$ in Theorem\ref{theorem.timeout1}, we know that~:
\[k = \sum_{i \leq T_{Generation}} \frac{n^{2} \cdot p_{a_{i}} \cdot p_{Neighbor}}{T \cdot N} (1 - E_{-}) \geq \]
\[ \geq \frac{1 - \lambda}{\lambda} \cdot timeout \cdot \frac{n^{2} \cdot p_{MAX} \cdot p_{Neighbor}}{T \cdot N} (1 - E_{-})\]

By selecting $timeout = m$ we verify that the information propagation process will be completed successfully, and so we can now write~:
\[timeout = \frac{2 (\rho -\ln(\frac{\epsilon}{n}) \big)}{1-e^{-\frac{k}{2n}}} \leq
\frac{2 \big(\rho -\ln(\frac{\epsilon}{n}) \big)}{1-e^{- \frac{1 - \lambda}{\lambda} \cdot timeout \cdot \frac{n \cdot p_{MAX} \cdot p_{Neighbor}}{2 T \cdot N} (1 - E_{-})}} \]

Using a similar approximation as in the previous section, we shall assume that the requested $\epsilon$ is polynomial in $\frac{1}{n}$. We can therefore write~:
\[2 (\rho + (\alpha + 1) \ln(n) \big) =
timeout (1-e^{- \frac{1 - \lambda}{\lambda} \cdot timeout \cdot \frac{n \cdot p_{MAX} \cdot p_{Neighbor}}{2 T \cdot N} (1 - E_{-}) })\]
Using the bound $(1 - x) < e^{-x}$ for $x < 1$ we can see that when assuming that~:
\[\frac{1 - \lambda}{\lambda} \cdot timeout \cdot \frac{n \cdot p_{MAX} \cdot p_{Neighbor}}{2 T \cdot N} (1 - E_{-}) < 1\]
we can write the previous expression as~:
\[2 \rho + (2\alpha + 2) \ln(n) \geq
\frac{1 - \lambda}{\lambda} \cdot timeout^{2} \cdot \frac{n \cdot p_{MAX} \cdot p_{Neighbor}}{2 T \cdot N} (1 - E_{-})\]
and therefore~:
\[ timeout \leq \sqrt{\frac{4 T \cdot N \left(\rho + (\alpha + 1) \ln(n) \right)}{n \cdot p_{MAX} \cdot p_{Neighbor} \cdot (1 - E_{-}) \frac{1 - \lambda}{\lambda}}} \]
which means that~:
\[ T_{Vac} \leq \frac{1}{\lambda} \sqrt{\frac{4 T \cdot N \left(\rho + (\alpha + 1) \ln(n) \right)}{n \cdot p_{MAX} \cdot p_{Neighbor} \cdot (1 - E_{-}) \frac{1 - \lambda}{\lambda}}} \]

In order to find the optimal value of $\lambda$ we now calculate $\frac{\partial T_{Vac}}{\partial \lambda}$~:
\[
\frac{\partial T_{Vac}}{\partial \lambda} =
- \frac{1 - 2\lambda}{2 (\lambda - \lambda^{2})^{1.5}} \sqrt{\frac{4 T \cdot N \left(\rho + (\alpha + 1) \ln(n) \right)}{n \cdot p_{MAX} \cdot p_{Neighbor} \cdot (1 - E_{-})}}
\]

It can now be easily seen that~:
\[
\lambda = \frac{1}{2} \quad \longrightarrow \quad \frac{\partial T_{Vac}}{\partial \lambda} = 0 \quad , \quad \frac{\partial^{2} T_{Vac}}{\partial \lambda^{2}} > 0
\]
and therefore the value of $\lambda = 0.5$ minimizes the upper bound over the algorithm's completion time.

\bibliographystyle{aaai}
\bibliography{/Altshuler}

\begin{thebibliography}{}

\bibitem[\protect\citeauthoryear{Adamic \bgroup et al.\egroup
  }{2001}]{probabilistic-flood-ttl1}
Adamic, L.~A.; Lukose, R.~M.; Puniyani, A.~R.; and Huberman, B.~A.
\newblock 2001.
\newblock Search in power-law networks.
\newblock {\em Phys. Rev. E} 64(4):046135.

\bibitem[\protect\citeauthoryear{Altshuler \bgroup et al.\egroup
  }{2008}]{UAV-ROBOTICA}
Altshuler, Y.; Yanovsky, V.; Bruckstein, A.; and Wagner, I.
\newblock 2008.
\newblock Efficient cooperative search of smart targets using uav swarms.
\newblock {\em ROBOTICA} 26:551--557.

\bibitem[\protect\citeauthoryear{Altshuler, Wagner, and
  Bruckstein}{2009}]{ICINCO2009}
Altshuler, Y.; Wagner, I.; and Bruckstein, A.
\newblock 2009.
\newblock Collaborative exploration in grid domains.
\newblock In {\em Sixth International Conference on Informatics in Control,
  Automation and Robotics (ICINCO)}.

\bibitem[\protect\citeauthoryear{Apap \bgroup et al.\egroup
  }{2002}]{apap-detecting-malicious}
Apap, F.; Honig, A.; Hershkop, S.; Eskin, E.; and Stolfo, S.
\newblock 2002.
\newblock {\em Recent Advances in Intrusion Detection}.
\newblock Springer Berlin Heidelberg.
\newblock chapter Detecting Malicious Software by Monitoring Anomalous Windows
  Registry Accesses,  36--53.

\bibitem[\protect\citeauthoryear{Bailey}{1975}]{epidemic_book}
Bailey, N.
\newblock 1975.
\newblock {\em The Mathematical Theory of Infectious Diseases and its
  Applications (second edition)}.
\newblock Hafner Press.

\bibitem[\protect\citeauthoryear{Barak \bgroup et al.\egroup
  }{2000}]{Faulty-Halevi}
Barak, B.; Halevi, S.; Herzberg, A.; and Naor, D.
\newblock 2000.
\newblock Clock synchronization with faults and recoveries (extended abstract).
\newblock In {\em PODC '00: Proceedings of the nineteenth annual ACM symposium
  on Principles of distributed computing},  133--142.
\newblock New York, NY, USA: ACM.

\bibitem[\protect\citeauthoryear{Broder \bgroup et al.\egroup
  }{1989}]{exploration_broder}
Broder, A.; Karlin, A.; Raghavan, P.; and Upfal, E.
\newblock 1989.
\newblock Trading space for time in undirected $s-t$ connectivity.
\newblock In {\em ACM Symposium on Theory of Computing (STOC)},  543--549.

\bibitem[\protect\citeauthoryear{Cagalj, Hubaux, and Enz}{2002}]{flooding_NP}
Cagalj, M.; Hubaux, J.; and Enz, C.
\newblock 2002.
\newblock Minimum-energy broadcast in all-wireless networks: Np-completness and
  distribution issues.
\newblock In {\em MOBICOM}.

\bibitem[\protect\citeauthoryear{Canalys}{2009}]{Canalys-smartphones}
Canalys.
\newblock 2009.
\newblock Canalys estimates, Canalys.

\bibitem[\protect\citeauthoryear{Castelluccia \bgroup et al.\egroup
  }{2006}]{Tsudik1}
Castelluccia, C.; Jarecki, S.; Kim, J.; and Tsudik, G.
\newblock 2006.
\newblock Secure acknowledgement aggregation and multisignatures with limited
  robustness.
\newblock {\em Computer Networks} 50(10):1639--1652.

\bibitem[\protect\citeauthoryear{Chakrabarti \bgroup et al.\egroup
  }{2008}]{Epidemic-Chakrabarti}
Chakrabarti, D.; Wang, Y.; Wang, C.; Leskovec, J.; and Faloutsos, C.
\newblock 2008.
\newblock Epidemic thresholds in real networks.
\newblock {\em ACM Trans. Inf. Syst. Secur.} 10(4):1--26.

\bibitem[\protect\citeauthoryear{Chung and Lu}{2001}]{ER_Diameter}
Chung, F., and Lu, L.
\newblock 2001.
\newblock The diameter of sparse random graphs.
\newblock {\em Advances in Applied Mathematics} 26:257--279.

\bibitem[\protect\citeauthoryear{Crisostomo, Barros, and
  Bettstetter}{2008}]{flooding_probability1}
Crisostomo, S.; Barros, J.; and Bettstetter, C.
\newblock 2008.
\newblock Flooding the network: Multipoint relays versus network coding.
\newblock In {\em 4th IEEE Intl. Conference on Circuits and Systems for
  Communications (ICCSC)},  119--124.

\bibitem[\protect\citeauthoryear{Demers \bgroup et al.\egroup
  }{1987}]{epidemic_basic1}
Demers, A.; Greene, D.; Hauser, C.; Irish, W.; Larson, J.; Shenker, S.;
  Sturgis, H.; Swinehart, D.; and Terry, D.
\newblock 1987.
\newblock Epidemic algorithms for replicated database maintenance.
\newblock In {\em In Proc. of the Sixth ACM Symp. on Principles of Distributed
  Computing},  1--12.

\bibitem[\protect\citeauthoryear{Dolev and Tzachar}{2010}]{dolev1}
Dolev, S., and Tzachar, N.
\newblock 2010.
\newblock Spanders: Distributed spanning expanders.
\newblock In {\em Proc. of the 25th ACM Symposium on Applied Computing
  (SAC-SCS)}.

\bibitem[\protect\citeauthoryear{Dolev, Schiller, and Welch}{2006}]{dolev2}
Dolev, S.; Schiller, E.; and Welch, J.
\newblock 2006.
\newblock Random walk for self-stabilizing group communication in ad hoc
  networks.
\newblock {\em IEEE Transactions on Mobile Computing} 5:893--905.

\bibitem[\protect\citeauthoryear{Eagle, Pentland, and
  Lazer}{2009}]{RealityMining}
Eagle, N.; Pentland, A.; and Lazer, D.
\newblock 2009.
\newblock Inferring social network structure using mobile phone data.
\newblock {\em Proceedings of the National Academy of Sciences (PNAS)}
  106:15274--15278.

\bibitem[\protect\citeauthoryear{Erdos and Renyi}{1959}]{ER_Graphs}
Erdos, P., and Renyi, A.
\newblock 1959.
\newblock On random graphs.
\newblock {\em Publ. Math. Debrecen} 6:290--291.

\bibitem[\protect\citeauthoryear{Erdos and Renyi}{1960}]{ER_Graphs2}
Erdos, P., and Renyi, A.
\newblock 1960.
\newblock On the evolution of random graphs.
\newblock {\em Publications of the Mathematical Institute of the Hungarian
  Academy of Sciences} 5:17--61.

\bibitem[\protect\citeauthoryear{Feige}{1995}]{Feige-random-upper1}
Feige, U.
\newblock 1995.
\newblock A tight upper bound on the cover time for random walks on graphs.
\newblock {\em Random Struct. Algorithms} 6(1):51--54.

\bibitem[\protect\citeauthoryear{Fragouli, Widmer, and
  Boudec}{2006}]{network_coding}
Fragouli, C.; Widmer, J.; and Boudec, J.~L.
\newblock 2006.
\newblock A network coding approach to energy efficient broadcasting: from
  theory to practice.
\newblock In {\em The 25th IEEE International Conference on Computer
  Communications (INFOCOM2006)},  1--11.

\bibitem[\protect\citeauthoryear{Ganesa \bgroup et al.\egroup
  }{2002}]{epidemic_neighborhood}
Ganesa, D.; Krishnamachari, B.; Woo, A.; Culler, D.; Estrin, D.; and Wicker, S.
\newblock 2002.
\newblock An empirical study of epidemic algorithms in large scale multihop
  wireless networks --- technical report ucla/csd-tr 02-0013.
\newblock Technical report, UCLA Computer Science.

\bibitem[\protect\citeauthoryear{Garetto, Gong, and
  Towsley}{2003}]{Garetto03modelingmalware}
Garetto, M.; Gong, W.; and Towsley, D.
\newblock 2003.
\newblock Modeling malware spreading dynamics.
\newblock In {\em In Proceedings of IEEE INFOCOM},  1869--1879.

\bibitem[\protect\citeauthoryear{Gemmel}{1997}]{TresholdCrypto1}
Gemmel, P.
\newblock 1997.
\newblock An introduction to threshold cryptography.
\newblock {\em CryptoBytes}  7--12.

\bibitem[\protect\citeauthoryear{GetJar}{2010}]{GetJar}
GetJar.
\newblock 2010.
\newblock Getjar statistics.

\bibitem[\protect\citeauthoryear{Hyponnen}{2006}]{viruses3-hyponnen}
Hyponnen, M.
\newblock 2006.
\newblock Malware goes mobile.
\newblock {\em Sci. American} 295:70.

\bibitem[\protect\citeauthoryear{Jonasson and
  Schramm}{2000}]{schramm-covertime-planar}
Jonasson, J., and Schramm, O.
\newblock 2000.
\newblock On the cover time of planar graphs.
\newblock {\em Electron. Comm. Probab.} 5:85--90 (electronic).

\bibitem[\protect\citeauthoryear{Kim, Smith, and Shin}{2008}]{Kim2008}
Kim, H.; Smith, J.; and Shin, K.~G.
\newblock 2008.
\newblock Detecting energy-greedy anomalies and mobile malware variants.
\newblock In {\em MobiSys '08: Proceeding of the 6th international conference
  on Mobile systems, applications, and services},  239--252.
\newblock New York, NY, USA: ACM.

\bibitem[\protect\citeauthoryear{Kleinberg}{2007}]{viruses1-kleinberg}
Kleinberg, J.
\newblock 2007.
\newblock The wireless epidemic.
\newblock {\em Nature} 449:287--288.

\bibitem[\protect\citeauthoryear{Koenig, Szymanski, and Liu}{2001}]{Koenig2}
Koenig, S.; Szymanski, B.; and Liu, Y.
\newblock 2001.
\newblock Efficient and inefficient ant coverage methods.
\newblock {\em Annals of Mathematics and Artificial Intelligence} 31:41--76.

\bibitem[\protect\citeauthoryear{Kong, Peng, and Rekleitis}{2006}]{rekleitis06}
Kong, C.; Peng, N.; and Rekleitis, I.
\newblock 2006.
\newblock Distributed coverage with multi-robot system.
\newblock In {\em IEEE International Conference on Robotics and Automation}.

\bibitem[\protect\citeauthoryear{Korf}{1990}]{LRTASTAR}
Korf, R.
\newblock 1990.
\newblock Real-time heuristic search.
\newblock {\em Artificial Intelligence} 42:189--211.

\bibitem[\protect\citeauthoryear{Liu \bgroup et al.\egroup }{2008}]{Voting-Liu}
Liu, D.; Ning, P.; Liu, A.; Wang, C.; and Du, W.~K.
\newblock 2008.
\newblock Attack-resistant location estimation in wireless sensor networks.
\newblock {\em ACM Trans. Inf. Syst. Secur.} 11(4):1--39.

\bibitem[\protect\citeauthoryear{Lv \bgroup et al.\egroup
  }{2002}]{probabilistic-flood-ttl2}
Lv, Q.; Cao, P.; Cohen, E.; Li, K.; and Shenker, S.
\newblock 2002.
\newblock Search and replication in unstructured peer-to-peer networks.
\newblock In {\em ICS '02: Proceedings of the 16th international conference on
  Supercomputing},  84--95.
\newblock New York, NY, USA: ACM.

\bibitem[\protect\citeauthoryear{mca}{2008}]{mcafee1}
2008.
\newblock Mcafee mobile security report 2008.
\newblock Technical report.

\bibitem[\protect\citeauthoryear{mca}{2009}]{mcafee2}
2009.
\newblock Mcafee mobile security report 2009.
\newblock Technical report.

\bibitem[\protect\citeauthoryear{Moskovitch \bgroup et al.\egroup
  }{2007a}]{elovici2-detection-malicious}
Moskovitch, R.; Gus, I.; Pluderman, S.; Stopel, D.; Glezer, C.; Shahar, Y.; and
  Elovici, Y.
\newblock 2007a.
\newblock Detection of unknown computer worms activity based on computer
  behavior using data mining.
\newblock In {\em CISDA 2007. IEEE Symposium on Computational Intelligence in
  Security and Defense Applications},  169--177.

\bibitem[\protect\citeauthoryear{Moskovitch \bgroup et al.\egroup
  }{2007b}]{elovici1-detection-malicious}
Moskovitch, R.; Pluderman, S.; Gus, I.; Stopel, D.; Feher, C.; Parmet, Y.;
  Shahar, Y.; and Elovici, Y.
\newblock 2007b.
\newblock Host based intrusion detection using machine learning.
\newblock In {\em 2007 IEEE Intelligence and Security Informatics},  107--114.

\bibitem[\protect\citeauthoryear{Mutz \bgroup et al.\egroup
  }{2006}]{Anomalous-Mutz}
Mutz, D.; Valeur, F.; Vigna, G.; and Kruegel, C.
\newblock 2006.
\newblock Anomalous system call detection.
\newblock {\em ACM Trans. Inf. Syst. Secur.} 9(1):61--93.

\bibitem[\protect\citeauthoryear{Narasimha, Tsudik, and
  Yi}{2003}]{TresholdCrypto3}
Narasimha, M.; Tsudik, G.; and Yi, J.~H.
\newblock 2003.
\newblock On the utility of distributed cryptography in p2p and manets: the
  case of membership control.
\newblock In {\em Proceedings of the 11th IEEE International Conference on
  Network Protocols},  336--345.

\bibitem[\protect\citeauthoryear{Ni \bgroup et al.\egroup
  }{1999}]{flooding_counterstorm}
Ni, S.; Tseng, Y.; Chen, Y.; and Sheu, J.
\newblock 1999.
\newblock The broadcast storm problem in a mobile ad hoc network.
\newblock In {\em In Proceedings of the ACM/IEEE International Conference on
  Mobile Computing and Networking (MOBICOM)},  151--162.

\bibitem[\protect\citeauthoryear{Qayyum and
  Laouiti}{2002}]{flooding_background5}
Qayyum, L., and Laouiti, A.
\newblock 2002.
\newblock Multipoint relaying for flooding broadcast messages in mobile
  wireless networks.
\newblock In {\em Proceedings of HICSS}.

\bibitem[\protect\citeauthoryear{Sasson, Cavin, and
  Schiper}{2003}]{flooding_background1}
Sasson, Y.; Cavin, D.; and Schiper, A.
\newblock 2003.
\newblock Probabilistic broadcas for flooding in wireless mobile ad-hoc
  networks.
\newblock In {\em Proceedings of IEEE Wireless communication and networks
  (WCNC)}.

\bibitem[\protect\citeauthoryear{Stojmenovic, Seddigh, and
  Zunic}{2002}]{flooding_knowledge2}
Stojmenovic, I.; Seddigh, M.; and Zunic, J.
\newblock 2002.
\newblock Dominating sets and neighbor elimination-based broadcasting
  algorithms in wireless networks.
\newblock {\em IEEE Transactions on Parallel and Distributed Systems}
  13(1):14--25.

\bibitem[\protect\citeauthoryear{Svennebring and Koenig}{2004}]{Svennebring1}
Svennebring, J., and Koenig, S.
\newblock 2004.
\newblock Building terrain-covering ant robots: A feasibility study.
\newblock {\em Autonomous Robots} 16(3):313--332.

\bibitem[\protect\citeauthoryear{van Renesse and
  Birman}{2002}]{epidemic_hierarchical}
van Renesse, R., and Birman, K.
\newblock 2002.
\newblock Scalable management and data mining using astrolabe.
\newblock In {\em In Proc. of the First International Workshop on Peer-to-Peer
  Systems (IPTPS'02)}.

\bibitem[\protect\citeauthoryear{Vogels, van Renesse, and
  Birman}{2003}]{epidemics-survey}
Vogels, W.; van Renesse, R.; and Birman, K.
\newblock 2003.
\newblock The power of epidemics: robust communication for large-scale
  distributed systems.
\newblock {\em SIGCOMM Comput. Commun. Rev.} 33(1):131--135.

\bibitem[\protect\citeauthoryear{Wagner \bgroup et al.\egroup }{2008}]{CC08}
Wagner, I.; Altshuler, Y.; Yanovski, V.; and Bruckstein, A.
\newblock 2008.
\newblock Cooperative cleaners: A study in ant robotics.
\newblock {\em The International Journal of Robotics Research (IJRR)}
  27(1):127--151.

\bibitem[\protect\citeauthoryear{Wang \bgroup et al.\egroup
  }{2009}]{barabasi-science09}
Wang, P.; Gonzalez, M.; Hidalgo, C.; and Barabasi, A.
\newblock 2009.
\newblock Understanding the spreading patterns of mobile phone viruses.
\newblock {\em Science} 324:1071--1075.

\bibitem[\protect\citeauthoryear{Williams and Camp}{2002}]{flooding_survey1}
Williams, B., and Camp, T.
\newblock 2002.
\newblock Comparison of broadcasting techniques for mobile ad hoc networks.
\newblock In {\em MOBIHOC},  9--11.

\bibitem[\protect\citeauthoryear{Zhou, Leckie, and
  Karunasekera}{2010}]{CollaborativeIntrusionDetection-2010}
Zhou, C.~V.; Leckie, C.; and Karunasekera, S.
\newblock 2010.
\newblock A survey of coordinated attacks and collaborative intrusion
  detection.
\newblock {\em Computers and Security} 29:124--140.

\end{thebibliography}
\end{document}